\documentclass[11pt]{article}
\usepackage{graphicx}
\usepackage{amssymb}
\usepackage{epsfig}
\usepackage{tikz}
\usepackage{pdfpages}
\usepackage[section]{placeins}

\usepackage[pagebackref]{hyperref}

\def\cadremath#1{\vbox{\hrule\hbox{\vrule\kern8pt\vbox{\kern8pt
			\hbox{ {$\displaystyle #1 $ } }\kern8pt} 
			\kern8pt\vrule}\hrule}}

\def\today{\number\day\space\ifcase\month\or Janvier \or F\'evrier \or  Mars
   \or Avril \or Mai \or Juin \or Juillet \or Ao\^ut \or Septembre \or Octobre
   \or Novembre \or D\'ecembre \fi\number \year}

\date{13 Decembre 2013}

\begin{document}

\title{Higher index focus-focus singularities in the Jayne-Cummings-Gaudin model :
symplectic invariants and monodromy.
}
\bigskip
\author{O. Babelon$^*$, B. Dou\c{c}ot\footnote{emails: babelon@lpthe.jussieu.fr, doucot@lpthe.jussieu.fr} \\
\hskip-2cm
Sorbonne Universit\'es, UPMC Univ Paris 06, UMR 7589, LPTHE, F-75005, Paris, France.
\footnote{Laboratoire de Physique Th\'eorique et Hautes Energies (LPTHE), Tour 13-14, 4\` eme \'etage, Boite 126,
4 Place Jussieu,  75252 Paris Cedex 05.}
\footnote{The LPTHE is an unit\'e mixte de recherche UPMC-CNRS and is part of the Institut Lagrange de Paris.}}
\bigskip

\maketitle  

\abstract{We study the symplectic geometry of the Jaynes-Cummings-Gaudin model with $n=2m-1$ spins. We show that there are focus-focus singularities of maximal Williamson type $(0,0,m)$. We construct the linearized normal flows in the vicinity of such a point and show that soliton type solutions extend them  globally on the critical torus. This allows us to compute the leading term in the Taylor expansion of the symplectic invariants  and the monodromy associated to this singularity.  }

\tableofcontents

\section{Introduction}

The theory of integrable systems started with the work of Liouville \cite{Liouville} where he defined the general concept of  integrable systems and 
their integration by ``quadratures''. Let ${\cal M}$ be a symplectic space of dimension $2n$, the system is Liouville integrable if we can define $n$ Hamiltonians  $H_i$ in involution such that ``generically'' $dH_1 \wedge dH_2 \wedge \cdots \wedge dH_n $ has maximal rank $r=n$. The remarkable Arnold-Liouville theorem  states that the space ${\cal M}$ is then fibered by  invariant $n$-dimensional Lagrangian tori. This is a {\em semi global} result as it gives a global information on the fiber, which is a torus, and it is sufficient to assert that the motion is quasi periodic almost everywhere. This fibration however contains singular fibers i.e. fibers containing points where the  rank $r$ is not maximal, which play a very important role for the global properties of the system, see e.g. \cite{Bolsinov06}. In this work, we will concentrate  on special singularities where the  rank $r=0$. They correspond to equilibrium points, which can be stable or unstable, and have also a very  rich physical content.

An equilibrium point $x^{(0)}$ is a simultaneous critical point for all the $H_i$, in other words, it is a point at which the differential of the moment
map $\mu$ vanishes. Expanding the $H_i$ around such a point, we get $n$ Poisson commuting quadratic forms $Q_i$. 

We shall consider the case of a purely focus-focus equilibrium point of Williamson type ($m_e=m_h=0$ and $m_{ff}=m$). 
In this case, the dimension of  $\mathcal{M}$ is equal to $4m$, and there exist canonical coordinates $(p_i,q_i)$
such that the quadratic forms $Q_i$ are linear combinations of the quadratic normal forms
\begin{eqnarray*}
K_j & = & p_{2j} q_{2j} + p_{2j+1}q_{2j+1}, \quad j=1,...,m \\
L_{j} & = & - p_{2j} q_{2j+1} + p_{2j+1} q_{2j}, \quad j=1,...,m
\end{eqnarray*}
By  Eliasson theorem~\cite{Eliasson90}, this description extends to a local neighborhood of the point $x_0$.   There exists 
a {\em symplectic} diffeomorphism $\Phi : U \subset \mathbb{R}^{2n} \longrightarrow \mathcal{V} \subset \mathcal{M}$ mapping the neighborhood $U$ of the origin 
to the neighborhood $\mathcal{V}$ of $x^{(0)}$, and a local diffeomorphism $\psi :  \mathbb{R}^{n} \longrightarrow \mathbb{R}^{n}$ in a neighborhood of the origin
such that:
$$
\mu \circ \Phi = \psi \circ \mu^{(0)} 
$$
where  $\mu^{(0)}$ is the quadratic moment map $(p,q)\to (K,L)$.
An important consequence of this theorem is that it allows to extend the flows associated to the quadratic generators
$K_j$ and $L_j$ from a neighborhood of $x^{(0)}$ ({\em local} level) to a family of fibers containing the level set of $x^{(0)}$ 
({\em semi global} level). These flows will be called {\em normal flows} and will play an important role in this paper. 

Using the complex variables
\begin{equation}
w_j =p_{2j}+i p_{2j+1}, \quad z_{j} = q_{2j} +i q_{2j+1}, \quad j=1,...,m
\label{wzpq}
\end{equation}
we have
$$
w_j\bar{z}_j=K_j + i L_j
$$
hence, the level set $\mathcal{S}$ of the equilibrium point is the image, by the diffeomorphism $\Phi$,  of the product of $m$ two dimensional 
components, themselves the union of two complex planes intersecting transversely
$$ 
{\cal C}_j : w_j =0 \;\mathrm{or}\; z_j =0,  \quad j=1,...,m
$$
It is a  $2m$-dimensional cone ${\cal C}$ 
$$ 
{\cal C} =\prod_{j=1}^{m} {\cal C}_j
$$

We see that  in a neighborhood of the equilibrium point, the phase-space fibration induced by the moment map 
is symplectically equivalent to the fibration of a direct product of $m$ independent integrable four dimensional dynamical systems with a focus-focus singularity.
An important question is to see whether such a simple description holds also at the semi-global level. 

In the Jaynes-Cummings-Gaudin model that we will consider in this paper,  the level set of the equilibrium point is a compact pinched torus of dimension $2m$. 
A result of Tien Zung~\cite{Zung96} asserts that, at the topological level,  this pinched torus  is  equivalent 
to a product of $m$ two dimensional pinched tori.
But this is not true at the symplectic level, and it is the purpose of this work to compute symplectic invariants defined in \cite{Dufour94,San03} preventing this simple product decomposition of the singular fiber.

\begin{figure}[h!]
\centering
\raisebox{3.4cm}{$T_{ff}=$}
\includegraphics[height= 7cm]{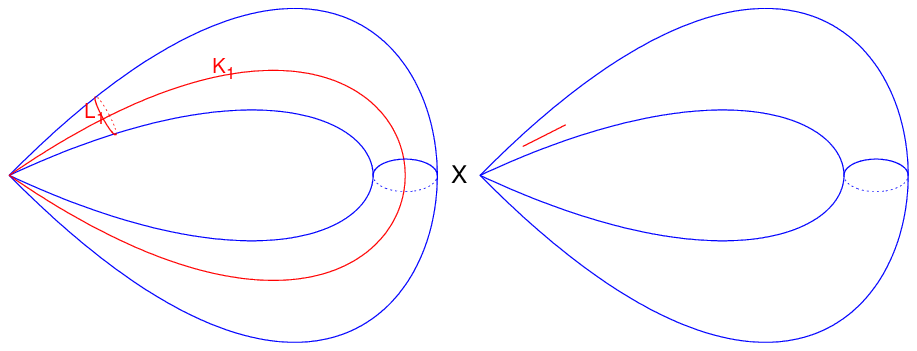}
 \raisebox{0cm}{${~~~~~~~~~~~~~~~~~~~~~~~~}^{X_1}~~~~~~~~~~~~~~~~~~~~~~~~~~~~~~~~~~~~~~~~~~~~~{~}^{X_k}$}  
\caption{The singular torus seen as a product of two dimensional pinched tori whose coordinates are the parameters $X_k$ entering the solitonic solution of the equations of motion. A big motion ($K_1$) on one of the components of the torus induces a  motion on the other components of the torus as well.}
\label{fig_decomposition}
\end{figure}

To achieve this goal, we will combine two informations. The first one is a very simple description of the  normal modes around the singularity, which 
provide a complete {\em local} description of the system. The second one, of a {\em semi global} nature, is provided by the explicit solitonic solutions of the equations of motion on the singular fibre.
It turns out that the two objects, normal modes and solitons, match perfectly~: solitons are just global extensions of normal modes to the full singular fiber. 
The $2m$ real parameters describing the initial conditions of the solitonic solution may be seen as coordinates on the singular fiber which precisely realize its decomposition into a product of $m$ two dimensional singular tori, thereby extending to the whole  fiber the decomposition provided by the normal coordinates in the vicinity of the critical point.

The plan of the paper is as follows. In Section \ref{section_definition} we recall the definition of the Jaynes-Cummings-Gaudin model and basic facts about its integrability. 
In Section \ref{section_normal_flows} we study the vicinity of an unstable point of Williamson type $(m_e=m_h=0,m_{ff}=m)$. 
In particular we  give a simple description of normal coordinates around that point. In Section \ref{section_normal_solitons} we relate these normal modes 
to solitonic solutions on the pinched torus, the level set of the unstable point. In Section \ref{comp_symplectic_invariants}  we construct closed periodic trajectories on Liouville tori close to the critical one. This requires  to take into account both diagonal \cite{San03} and non diagonal  \cite{Dullin07}  components of the motion which signal the obstruction  to Tien Zung decomposition at the symplectic level.
The action variables associated to these periodic motion provide  the symplectic invariants 
in the sense of San V\~{u} Ngoc. We  use 
them to determine Duistermaat's monodromy \cite{Duistermaat80}. A number of technical details are provided in the Appendices for the sake of completeness.

\bigskip

{\bf Acknowlegment.} We wish to thank San V\~u Ngoc for very helpful discussions.

\section{The Jaynes-Cummings-Gaudin model}
\label{section_definition}
\subsection{Basic definitions and integrability}

This model describes a collection of $n$ spins  coupled to a single harmonic oscillator.
It derives from the  Hamiltonian:
\begin{equation}
H= \sum_{j=1}^{n} (2 \epsilon_j +\omega)  s_j^z  + \omega \bar{b} b + 
\sum_{j=1}^{n} \left( \bar{b} s_j^-+ b s_j^+ \right)
\label{bfconspinclas}
\end{equation}
The $\vec{s}_j$ are spin variables, and $b,\bar{b}$ is a  harmonic oscillator.
The Poisson brackets read
\begin{equation}
\{ s_j^a , s_j^b \} = - \epsilon_{abc} s_j^c, \quad \{ b , \bar{b} \} = i
\label{poisson}
\end{equation}
The $\vec{s}_j$ brackets are degenerate. We fix the value of the Casimir functions
$$
 || \vec{s}_j||^2 = \sum_{a=1}^3 s_j^a s_j^a = s^2
$$
Phase space has dimension $2(n+1)$. In the Hamiltonian we have used
$s_j^\pm = s_j^1 \pm i s_j^2$
which have Poisson brackets $ \{ s_j^z , s_j^\pm \} = \pm i s_j^\pm,\quad \{s_j^+,s_j^-\} = 2 i s_j^z$. 
The equations of motion read
\begin{eqnarray}
\dot{b} &=&  -i \omega b - i  \sum_{j=1}^{n} s_j^-  \label{motionb} \\
\dot{s_j^z} &=&  i  ( \bar{b} s_j^- - b s_j^+ ) \label{motionsz}\\
\dot{s_j^+} &=& i(2 \epsilon_j+\omega) s_j^+ -2i \bar{b} s_j^z \label{motions+} \\
\dot{s}_j^- &=&  -i(2 \epsilon_j +\omega) s_j^- +2i b s_j^z \label{motions-}
\end{eqnarray}
Introducing the Lax matrices
\begin{eqnarray}
L(\lambda) &=& 2 \lambda \sigma^z + 2 (b \sigma^+ + \bar{b} \sigma^-) + \sum_{j=1}^{n} {\vec{s}_j \cdot \vec{\sigma} \over \lambda-\epsilon_j} 
\label{defLaxL} \\
M(\lambda) &=& -i\lambda \sigma^z  -i{ \omega\over 2}\sigma^z 
-i (b\sigma^+ + \bar{b} \sigma^-)
\label{defLaxM}
\end{eqnarray}
where $\sigma^a$ are the Pauli matrices, $\sigma^\pm = {1\over 2} (\sigma^x \pm i \sigma^y)$,
$$
\sigma^x = \pmatrix{ 0 & 1 \cr 1 & 0}, \quad \sigma^y = \pmatrix{ 0 & -i \cr i & 0},\quad  \sigma^z = \pmatrix{1 & 0 \cr 0 & -1},\quad 
$$
it is not difficult to check that the equations of motion are equivalent to the Lax equation
\begin{equation}
\dot{L}(\lambda) = [M(\lambda), L(\lambda) ]
\label{eqLax}
\end{equation}
Letting
$$
L(\lambda) = \pmatrix{ A(\lambda) & B(\lambda) \cr C(\lambda) & -A(\lambda) }
$$
we have
\begin{eqnarray}
A(\lambda) &=& 2\lambda+ \sum_{j=1}^{n} {s_j^z \over \lambda - \epsilon_j } \label{defA}\\
B(\lambda) &=& 2b + \sum_{j=1}^{n} {s_j^- \over \lambda - \epsilon_j } \label{defB} \\
C(\lambda) &=& 2\bar{b} + \sum_{j=1}^{n} {s_j^+ \over \lambda - \epsilon_j }  \label{defC}
\end{eqnarray}
The  non vanishing Poisson brackets of these functions are simple :
\begin{eqnarray}
\{A(\lambda), B(\mu) \} &=& {i\over \lambda - \mu} ( B(\lambda) - B(\mu) ) \label{AB}\\
\{A(\lambda), C(\mu) \} &=& -{i\over \lambda - \mu} ( C(\lambda) - C(\mu) ) \label{AC}\\
\{B(\lambda), C(\mu) \} &=& {2i\over \lambda - \mu} ( A(\lambda) - A(\mu) ) \label{BC}
\end{eqnarray}
It follows  immediately that $ \rm{Tr}\,(L^2(\lambda) ) = 2 A^2(\lambda) + 2 B(\lambda) C(\lambda)  $ Poisson commute for different values of the spectral parameter:
$$
\{ \rm{Tr}\,(L^2(\lambda_1) ), \rm{Tr}\,(L^2(\lambda_2) ) \} = 0
$$
Hence $\Lambda(\lambda)\equiv \frac{1}{2}\rm{Tr}\,(L^2(\lambda) ) $ generates Poisson commuting quantities.   One has
\begin{eqnarray}
\Lambda(\lambda)  =  {Q_{2n+2}(\lambda)\over \prod_j (\lambda-\epsilon_j)^2}=  4\lambda^2  + 4  H_{n+1}  + 
 2 \sum_{j=1}^{n}  
{H_j  \over \lambda - \epsilon_j}  +  \sum_{j=1}^{n} {s^2 \over ( \lambda - \epsilon_j)^2 }
\label{detL}
\end{eqnarray}
where the $(n+1)$ commuting Hamiltonians $H_j$, $j=1,\cdots , n+1$  read
\begin{equation}
H_{n+1} =  b\bar{b} +  \sum_j s_j^z 
\label{Hn}
\end{equation}
and
\begin{equation}
H_j =  2\epsilon_j  s_j^z +  ( b s_j^+ + \bar{b} s_j^-)
+ \sum_{k\neq j} {s_j \cdot s_k \over \epsilon_j - \epsilon_k }, \quad j=1,\cdots , n
\label{Hj}
\end{equation}
The physically interesting Hamiltonian eq.(\ref{bfconspinclas}) is
\begin{equation}
H = \omega H_{n+1} + \sum_{j=1}^{n} H_j
\label{hamphys}
\end{equation}

\subsection{Critical points}

The critical points are equilibrium points for all the Hamiltonians $H_j$, $j=1,\cdots, n+1$. At such points the derivatives with respect of all coordinates on phase space vanish. 
In particular, since
$$
{\partial H_{n+1}\over \partial \bar{b}} = b,\quad {\partial H_j\over \partial \bar{b}} =  s_j^-
$$
we see that the critical points must be located at
\begin{equation}
b=\bar{b} =0, \quad s_j^{\pm} = 0, \quad s_j^z = e_j s, \quad e_j = \pm 1
\label{static}
\end{equation}
When we expand around a configuration eq.(\ref{static}), all the quantities ($b$, $\bar{b}$, $s_j^+$, $s_j^-$) are first order, but $s_j^z$ is second order because
$$
s_j^z = e_j \sqrt{s^2-s_j^+ s_j^- } = s e_j -{e_j\over 2s} s_j^+ s_j^-  + \cdots, \quad e_j = \pm 1
$$
It is then simple to see that all first order terms in the expansions of the Hamiltonians $H_j$ vanish. Hence we have found  $2^n$ critical points.

\section{The vicinity of an unstable fixed point}
\label{section_normal_flows}
\subsection{Normal Forms}

In the following discussion, we are considering the system with an odd number of spins $n=2m-1$ with $m \geq 1$, and we are assuming that
the equilibrium point defined by eq.~(\ref{static}) has the Williamson type $m_e=m_h=0$ and $m_{ff}=m$, i.e. it consists locally of a direct
product of $m$ elementary focus-focus singularities. It has been shown before that such points can be obtained by choosing the Zeeman energies
$\epsilon_j$ in a suitable region in parameter space~\cite{BD11,Yuzbashyan08}.   The quadratic normal form around an equilibrium point can be easily
obtained from the Lax representation~\cite{Krich83, BD11} and this is recalled in Section \ref{sec_Normal_Forms} below. The above assumption on the Williamson
type is equivalent to the statement  that the classical Bethe equation~(\ref{classicalBethe}) has no real root (hence the choice $n$ odd). Since the equation is real, it has  $m$ pairs of complex conjugated roots $E_{j},\bar{E}_{j}$, for $1\leq j \leq m$.
To first order in small deviations around the equilibrium point, the normal coordinates are  
$B(E_j)$, $B(\bar{E}_j)$, $C(E_j)$, and $C(\bar{E}_j)$ where the functions $B(\lambda), C(\lambda)$ are those defined in eqs.(\ref{defB},\ref{defC}).
The precise relation with the  canonical coordinates eq.(\ref{wzpq}) is:
\begin{eqnarray}
B(E_j) &=&  -i a'(E_j) z_j, \quad B(\bar{E}_j) =  w_j  \label{normal_coordinates_a} \\
C(E_j) &=&  \bar{w}_j,  \quad \quad \quad \quad ~~
C(\bar{E}_j) =  ia'(\bar{E}_j) \bar{z}_j  
\label{normal_coordinates_b}
\end{eqnarray}
These expressions for $z_j$ and $w_j$ as  functions of the dynamical variables of the Gaudin model 
should be understood as giving the differential of the function $\Phi^{-1}$ at the equilibrium point.  
An important consequence of Eliasson's theorem is that the knowledge of this differential is sufficient
to identify the generators of the normal flows, as long as we are interested in their restriction to the
singular torus.   

The symplectic form in terms of normal coordinates reads:
$$
\omega^{(0)} = {1\over 2} \sum_j dz_j \wedge d\bar{w}_j + d \bar{z}_j \wedge d w_j
$$
As explained in the Introduction, the components of the moment map $\mu^{(0)}$ are the quadratic Hamiltonians:
$$
L_j= {1\over 2i} (\bar{z}_j w_j -  z_j \bar{w}_j), \quad K_j={1\over 2} (\bar{z}_j w_j +  z_j \bar{w}_j)
$$
The Hamilton equations of motion associated to $\omega^{(0)}$ are:
$$
\dot{z}_j = 2{\partial H \over \partial \bar{w}_j},\quad \dot{w}_j = -2{\partial H \over \partial \bar{z}_j},\quad
\dot{\bar{z}}_j = 2{\partial H \over \partial w_j}, \quad \dot{\bar{w}}_j = -2{\partial H \over \partial z_j}
$$
From them, we see that the flows associated to $K_j, L_j$ are
$$
K_j:   (z_j \to e^{a_j} z_j,  w_j \to e^{-a_j} w_j), \quad L_j:  (z_j \to e^{i\theta_j} z_j , w_j \to e^{i\theta_j} w_j)
$$

\subsection{Generators of normal flows}

We begin by expanding the function $\mu^{(0)}\circ \Phi^{-1}$ to second order in small deviations from
the equilibrium point, recalling that $B(E_j)$, $B(\bar{E}_j)$, $C(E_j)$, and $C(\bar{E}_j)$ are first
order in these deviations. This gives:
\begin{eqnarray}
K_j &=& {i\over 2}\left( {B(E_j) C(E_j) \over a'(E_j)} -  {B(\bar{E}_j) C(\bar{E}_j) \over a'(\bar{E}_j)} \right) 
\label{normalH}\\ 
L_j &=& -{1\over 2}\left( {B(E_j) C(E_j) \over a'(E_j)} +  {B(\bar{E}_j) C(\bar{E}_j) \over a'(\bar{E}_j)} \right)
\label{normalL}
\end{eqnarray}
But Eliasson's theorem implies that $\mu^{(0)}\circ \Phi^{-1}= \psi^{-1} \circ \mu$, so we should be able to express
these quantities in terms of the conserved Hamiltonians $H_1$,...,$H_{n+1}$. In Section~\ref{sec_Normal_Forms} below,
we show that:
\begin{equation}
B( E_j) C( E_j) = 4 E_j^2 + 4 H_{n+1} + \sum_{i=1}^n {2H_i\over  E_j-\epsilon_i} + \sum_{k=1}^n {s^2\over ( E_j-\epsilon_i)^2} + ...  
\label{first_order_psi_inv}
\end{equation}
where the neglected terms are of order 3 and higher in small deviations from the equilibrium point.
This formula enables to construct the Taylor expansion of the function $\psi^{-1}$ up to first order. 
We get:
\begin{eqnarray}
K_{_j} &=& i\left( {2\over a'(E_j)} -  {2\over a'(\bar{E}_{j})} \right) \delta H_{n+1} + i \sum_{i=1}^n \left( {1\over a'(E_j)(E_j-\epsilon_i)} -  
{1\over a'(\bar{E}_{j}) (\bar{E}_{j} - \epsilon_i)} \right) \delta H_i \label{normal_generator_K} + ... \label{Kjquad}\\
L_{_j} &=& -\left( {2\over a'(E_j)} +  {2\over a'(\bar{E}_{j})} \right) \delta H_{n+1} - \sum_{i=1}^n \left( {1\over a'(E_j)(E_j-\epsilon_i)} +  
{1\over a'(\bar{E}_{j}) (\bar{E}_{j} - \epsilon_i)} \right) \delta H_i \label{normal_generator_L} + ...\label{Ljquad}
\end{eqnarray}
Here $\delta H_i= H_i - H_{i,\mathrm{eq}}$, where $H_{i,\mathrm{eq}}$ denotes the value of $H_i$ at the equilibrium point.
The neglected term are of order at least two in the $\delta H_i$'s. When we consider fibers of the moment map closer and closer
to the fiber containg the equilibrium point, the flows generated by the linear terms in the $\delta H_i$'s in
equations~(\ref{normal_generator_K}) and (\ref{normal_generator_L}) tend towards the normal flows defined in the Introduction. 
In particular, they coincide with these normal flows on  the fiber containing the equilibrium point.

\section{Normal flows on a singular torus}
\label{section_normal_solitons}
\subsection{Normal flows and soliton solutions}

From the previous discussion, the normal flows on the singular torus are given by the Hamiltonian flows
associated to the generators $K_j$ and $L_j$ defined in terms of the conserved Hamiltonians in eqs.~(\ref{normal_generator_K}) and (\ref{normal_generator_L}).
On the singular torus, these flows take the form of solitonic trajectories.
For the reader's convenience a complete derivation of the soliton formulae, which were constructed in ~\cite{Yuzbashyan08}, 
is given in Section~\ref{sec_n+1_degenerate} below. Here, we summarize the main
features of these solitonic solutions.  

We recall that $n=2m-1$,  and we assume that $ {\cal E}^0 = \emptyset$ in the construction of Section \ref{sec_n+1_degenerate}.
We  define  column vectors $(E^j)_l = E_l^j$  and $(XE^j)_l = X_lE_l^j$, $l=1 \cdots 2m$,  and two polynomials 
$$
{\cal P}^-(\lambda) = {1\over D_{n+1}} {\rm det} \pmatrix{ 1 & \lambda & \cdots & \lambda^{n_-} & 0 & 0     & \cdots &  0 \cr
                                                                                                      1 & E            & \cdots & E^{n_-}             & X & X E & \cdots & XE^{n_+} }
$$
and
$$
{\cal P}^+(\lambda) = {1\over D_{0}} {\rm det} \pmatrix{ 0 & 0 & \cdots & 0             & 1  & \lambda     & \cdots &  \lambda^{n_+} \cr
                                                                                                  1 & E  & \cdots & E^{n_-} & X & X E             & \cdots  & XE^{n_+ } }
$$
where $D_0$ and $D_{n+1}$ are the determinants:
\begin{eqnarray}
D_0&=& {\rm det} \pmatrix{ 1 & E & \cdots & E^{n_-} & X& XE& \cdots &  XE^{n_+-1}}  \label{D0} \\
D_{n+1}&=&  {\rm det} \pmatrix{ 1& E& \cdots& E^{n_- -1}& X& XE& \cdots &  XE^{n_+}}  \label{Dn+1}
\end{eqnarray}
The indices $n_\pm$ are defined as
$$
n_+={\rm degree}({\cal P}^+(\lambda)  ) =  {1\over 2} (n-1)=m-1, \quad n_- = {\rm degree}({\cal P}^-(\lambda)  ) =  {1\over 2} (n+1)=m
$$
The $X_l$ are given by:
\begin{equation}
 X_l(\{ t \}) = X_l(0) e^{\sqrt{-1}\left( \sum_i {s e_i\over E_l-\epsilon_i} t_i - t_{n+1}\right)},
\label{solitonmultitemps}
\end{equation}
where $t_i$ are the times flows associated to the Hamiltonians $H_i$ and $e_i=\pm 1$ characterize the critical point $s_i^z = s e_i$. The constants 
$X_l(0)$ are subjected to the reality conditions:
\begin{equation}
\overline{X_l(\{ t \})} X_{\bar{l}}(\{ t \}) =
\overline{X_l(0)} X_{\bar{l}}(0) = -{1\over 4} 
\label{real_cond}
\end{equation}
where $X_{\bar{l}}$ is associated to $E_{\bar{l}} \equiv \overline{E_l}$.
The function $C(\lambda)$ is given by :
\begin{eqnarray}
C(\lambda) &=&  \label{Clambda}\\
&&\hskip -1.5cm
2 {  {\rm det} \pmatrix{ 1 & \lambda & \cdots & \lambda^{n_-} & 0 & 0     & \cdots &  0 \cr
                                                                                                      1 & E            & \cdots & E^{n_-}             & X & X E & \cdots & XE^{n_+} }
 {\rm det} \pmatrix{ 0 & 0 & \cdots & 0             & 1  & \lambda     & \cdots &  \lambda^{n_+} \cr
                                                                                                  1 & E  & \cdots & E^{n_-} & X & X E             & \cdots  & XE^{n_+ } }
\over  \prod_j (\lambda - \epsilon_j) \;\;  {\rm det} \pmatrix{  1& E& \cdots& E^{n_- -1}& X& XE& \cdots &  XE^{n_+} }^2 }       
\nonumber                                                                                              
\end{eqnarray}

Let us now specialize the evolution equation~(\ref{solitonmultitemps}) to the normal flows defined through their 
generators~(\ref{normal_generator_K}) and (\ref{normal_generator_L}). Let us denote by $a_l$ (resp. $\theta_l$)
the time variable of the normal flow generated by $K_l$ (resp. $L_l$).
For an observable $X(t_1,\cdots , t_{n+1})$, we have
\begin{eqnarray*}
X(a_l) &=& X\left( t_i = i \left( {1\over a'(E_l)(E_l-\epsilon_i)} -  
{1\over a'(\bar{E}_{l}) (\bar{E}_{l} - \epsilon_i)} \right) a_l,\;\; t_{n+1} = i \left( {2\over a'(E_l)} -  {2\over a'(\bar{E}_{l})} \right) a_l \right)\\
X(\theta_l) &=& X\left( t_i =  -\left( {1\over a'(E_l)(E_l-\epsilon_i)} +  
{1\over a'(\bar{E}_{l}) (\bar{E}_{l} - \epsilon_i)} \right) \theta_l,\;\; t_{n+1} = - \left( {2\over a'(E_l)} +  {2\over a'(\bar{E}_{l})} \right) \theta_l \right) 
\end{eqnarray*}
Let us apply this formula to $X_k(\{t\})$ in eq.(\ref{solitonmultitemps}). We have to evaluate:
$$
 {1\over a'(E_l) }\sum_{i=1}^n {s e_i \over (E_k-\epsilon_i) (E_l-\epsilon_i)}
$$
Suppose first $k\neq l$. We have:
\begin{eqnarray*}
{1\over a'(E_l) }\sum_{i=1}^n {s e_i \over (E_k-\epsilon_i) (E_l-\epsilon_i)} &=&  {1\over a'(E_l) }\sum_{i=1}^n 
{1\over E_k-E_l} \left( {se_i\over E_l-\epsilon_i}-{se_i\over E_k-\epsilon_i} \right) \\
&=&  {1\over a'(E_l) }\sum_{i=1}^n {1\over E_k-E_l}  (-2E_l +2 E_k) = {2\over a'(E_l) }
\end{eqnarray*}
where we have used the classical Bethe equation:
$$
a(E) = 2E + \sum_i {se_i\over E-\epsilon_i} =0
$$
If $k=l$, we have directly:
$$
{1\over a'(E_l) }\sum_{i=1}^n {s e_i \over (E_l-\epsilon_i)^2} =  {1\over a'(E_l) } (2-a'(E_l) )
$$
Using these sums in  eq.(\ref{solitonmultitemps}), we arrive at the very simple result:
\begin{equation}
X_k(a_l) = X_k(0) e^{a_l (\delta_{kl} - \delta_{k\bar{l}})} , \quad 
X_k(\theta_l) = X_k(0) e^{i \theta_l (\delta_{kl} + \delta_{k\bar{l}})} 
\label{X_normal_flow}
\end{equation}
It is clear that the reality condition~(\ref{real_cond}) is preserved by these flows.  
The meaning of this formula is that the solitons can be viewed as a non-linear global extension of the normal modes on the critical torus. 
The $a_l$ trajectories give $m$ non compact cycles, while the $\theta_l$ trajectories give the remaining $m$ compact cycles.  In view 
of eq.(\ref{solitonmultitemps}) we see that the soliton solution depends on $2m$ complex parameters $X_l(0)$ submitted to the $m$ complex conditions 
eq.(\ref{real_cond}). Hence there remains $m$ complex (or $2m$ real) parameters which may be viewed as coordinates on the singular torus. 
These solitonic coordinates can be interpreted as generalizing at the semi global level the decomposition of the singular fiber into a product of two dimensional singular tori that is performed by normal coordinates in the vicinity of the critical point.

\subsection{Solitonic picture of the tangent cone}

We would like now to precise the connection between these solitonic formulae and the linearized flows in the vicinity of the equilibrium point.
As discussed in the Introduction, the tangent cone to the singular torus at the equilibrium point is the  product of $m$ two dimensional components ${\cal C}_k$.
There is an important relation between the  components ${\cal C}_k$ of the cone ${\cal C}$ and the flows  associated to the Hamiltonians $K_j$ on the critical manifold. 
In normal coordinates the orbits of generators $L_j$ are closed circles, which become closed orbits when mapped by $\Phi$ on phase-space $\mathcal{M}$.
The $m$ other generators $K_j$ induce flows which become unbounded when $a_j$ goes to $\pm \infty$ and the trajectories eventually leave  the open neighborhood $U$ in which the diffeomorphism $\Phi$ is defined. Notice however that when $a_j\to + \infty$,  $z_j=0$ is a contracting manifold for the flow $K_j$   and $w_j=0$ is an expanding  manifold. When $a_j\to - \infty$ the situation is opposite. Hence  a pattern ${\cal S}$ of $m$ signs $a_j \to \pm \infty$ identifies  a unique contracting manifold  ${\cal C}_{\cal S}^{contracting}$. 

Because there is only one critical point on the critical manifold, when we send  all the $a_j$ to $\pm\infty$ we  end up into a neighborhood of the critical point. 
The contracting manifold ${\cal C}_{\cal S}^{contracting}$ is therefore the asymptotic manifold when  all the  $a_j$'s are sent to infinity with the associated sign pattern.

\begin{figure}[h!]
\centering
\includegraphics[height= 5cm]{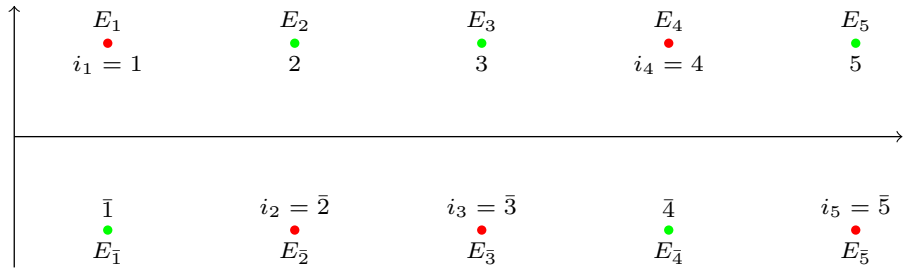}
\caption{The red dots represent the subset ${\cal I}= \{i_1=1, i_2= \bar{2}, i_3=\bar{3} ,i_4=4,i_5=\bar{5} \}$. The green dots represent the 
complementary (and conjugate) subset $\bar{\cal I} = \{\bar{i_1}=\bar{1}, \bar{i_2}= 2, \bar{i_3}=3 ,\bar{i_4}=\bar{4},\bar{i_5}=5 \}$.
}
\label{ibar}
\end{figure}

From eq.~(\ref{X_normal_flow}), the large $a_j$ limit corresponds to:
$$
|X_j| \rightarrow \infty \; \mathrm{and} \;  |X_{\bar{j}} | \rightarrow 0 \;\;\mathrm{if} \; a_j\to + \infty
$$
$$
|X_{\bar{j}}| \rightarrow \infty \; \mathrm{and} \; |X_{j}| \rightarrow 0 \;\;\mathrm{if} \; a_j\to - \infty
$$
To connect the soliton solutions with the definition of the tangent cone, we have to take this limit in the soliton formulae.
Let ${\cal I}$ the set of $m$ indices $(i_1,i_2,\cdots i_m) \subset \{1\cdots m, \bar{1} \cdots \bar{m}\}$ such that $X_{i_j} \to \infty$, specifying a contracting component ${\cal C}_{\cal S (\cal I)}^{contracting}$ of the tangent cone. 
If $j\in {\cal I}$ we still denote by $\bar{j}$ the label of the complex conjugate root $ \overline{E_j} \equiv E_{\bar{j}}$ (so that $\bar{\bar{j}}=j$).
Then   because of the reality condition eq.(\ref{real_cond}), $\bar{j} \notin {\cal I}$ and the indices $\bar{j}$ belong to the complementary set $\overline{\cal I}$ of ${\cal I}$ in $\{1\cdots m, \bar{1} \cdots \bar{m}\}$ (see Fig. \ref{ibar}).

It is useful to relate this definition of the components ${\cal C}_{\cal S (\cal I)}^{contracting}$ of the tangent cone with the one based on the normal
coordinates introduced in Section~\ref{section_normal_flows}. As we have said, either $j \in {\cal I}$ or $\bar{j} \in {\cal I}$.
If $j \in {\cal I}$, $z_j \rightarrow 0$, so according to eqs.~(\ref{normal_coordinates_a}),(\ref{normal_coordinates_b}),
$|C(E_{\bar{j}})| << |C(E_j)|$. If $\bar{j} \in {\cal I}$, the
converse holds and $|C(E_j)| << |C(E_{\bar{j}})|$. To check that this is indeed true, we need to know the asymptotic expression
for $C(\lambda)$ when $|X_j| \rightarrow \infty$ for any $j \in {\cal I}$. Using  eq.~(\ref{Clambda}) and 
the asymptotic expressions eqs.~(\ref{D1_approx}), (\ref{D2_approx}), (\ref{D_approx})
given in Section~\ref{asymptotic} below, we see that on the component ${\cal C}_{\cal I}$ of tangent cone, $C(\lambda)$ behaves as:
\begin{equation}
C(\lambda)\simeq \frac{2}{\prod_i (\lambda - \epsilon_i)}\prod_{i \in {\cal I} }(\lambda-E_{\bar{i}})
\sum_{j\in {\cal I}}\left(\frac{E_{\bar{j}}-E_j}{X_j}\prod_{k\in {\cal I} \atop k \neq j }\frac{E_{\bar{k}}-E_j}{E_k-E_j} (\lambda-E_k)\right)
\label{Clambda_approx}
\end{equation}
This expression shows that $C(\lambda)$ goes to zero when all the  $X_j$'s go to infinity, for $j \in {\cal I}$. This shows
that we are indeed describing a $2m$-dimensional subspace (over $\mathbb{R}$) of the tangent space at the equilibrium point.
Next, we see that, at this leading order, $C(\lambda)$ vanishes at $\lambda=E_{\bar{j}}$. 
In reality, $C(E_{\bar{j}})$ is not exactly zero, because of the subleading terms included in the determinant $D_1$ defined in
Section~\ref{asymptotic}. If the typical value of the $X_j$'s scales like $X$, the ratio of these subleading terms over the
dominant one scales like $X^{-2}$. From this, we conclude that $C(E_{\bar{j}})/C(E_j) \rightarrow 0$ when $X \rightarrow \infty$,
as expected.
More precisely, if $j\in {\cal I}$, to leading order in the scale $X$, we have from eq.~(\ref{Clambda_approx}): 
\begin{equation}
C(E_j, X_{i \in {\cal I}}  \to \infty) \simeq - \frac{2}{X_j \prod_i (E_j - \epsilon_i)} \prod_{k\in {\cal I}}(E_j-E_{\bar{k}})^{2}, \quad C(E_{\bar{j}}) =O(X^{-3})
\label{C_contracting}
\end{equation}

In the  limit  $a_j\to -\infty$, the dominant variables become the $X_{\bar{j}}$'s. The leading behavior
for $C(\lambda)$ is now given by eq.~(\ref{Clambda_approx}) where the set ${\cal I}$ is replaced by the complementary set $\overline{\cal I}$ (in
which all indices are replaced by $i\to \bar{i}$).
The state of the system is now labelled by the coordinates $B(E_j)=\overline{C(E_{\bar{j}})}$.
Using the reality condition~(\ref{real_cond}) and replacing $E_j$'s by $E_{\bar{j}}$'s in eq.~(\ref{C_contracting}) gives for $j\in {\cal I}$:
\begin{equation}
B(E_j,X_{i \in {\cal I}}  \to 0) \simeq \frac{8X_j}{\prod_i (E_j - \epsilon_i)} \prod_{k\in {\cal I}}(E_j-E_{\bar{k}})^{2}, \quad B(E_{\bar{j}})=O(X^3)
\label{B_expanding}
\end{equation}
where the scale $X$ goes to zero.

\section{ Periodic flows, symplectic invariants and monodromy}
\label{comp_symplectic_invariants}

\subsection{Motivation}

\begin{figure}[h!]
\centering
\includegraphics[height= 5cm]{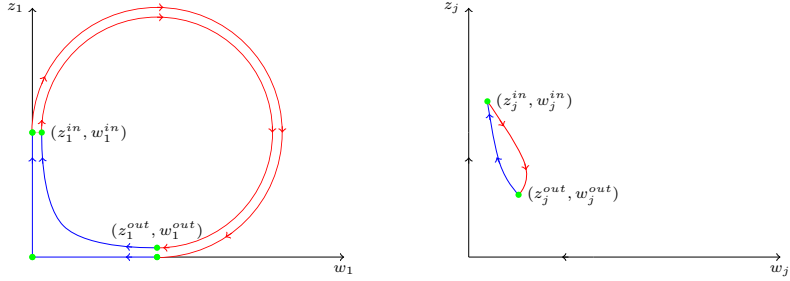}
\caption{The red curves represent the large motion of the $\alpha_1 K_1 + \beta_1 L_1$ flow projected on the $z_1,w_1$ manifold in the soliton case (outer curve) and 
on a nearby Liouville torus (inner curve). The motion starts from a position $(z_1^{in},w_1^{in}\simeq 0)$ on the expanding manifold and returns to a position 
$(z_1^{out}\simeq 0,w_1^{out})$ on the contracting manifold. At the same time on the $j^{th} >1$ component, there is an induced motion  from $(z_j^{in},w_j^{in})$ to
$(z_j^{out},w_j^{out})$. One can close the trajectory by adding an extra  motion $\alpha_1 K_1 + \beta_1 L_1$ from $(z_1^{out} ,w_1^{out})$ to $(z_1^{in},w_1^{in})$ and   a motion $\alpha_j K_j + \beta_j L_j$ from $(z_j^{out},w_j^{out})$ to $(z_j^{in},w_j^{in})$ (blue segments). These extra motions lie entirely in the domain of validity of the normal form.
}
\label{rangeS}
\end{figure}

In an integrable system, the action integrals $S_{\gamma}=\int_{\gamma}\alpha$, where the symplectic form $\omega = d\alpha$   and 
 $\gamma$ denotes any closed
cycle on regular Arnold--Liouville tori, are smooth functions of the conserved quantities in any open set 
containing only regular values of the moment map. These functions are also the generators of periodic flows
on Arnold-Liouville tori. Inspired by previous works~\cite{Dufour94,San03,Dullin07}, we wish to examine the
construction of periodic flows in the vicinity of a singular torus. We know already that $m$ such flows are
generated by the $L_j$'s. As we have seen, on the singular torus, the flow generated by $K_j$ corresponds to a
large motion, moving away from the critical point $x^{(0)}$ along the corresponding expanding manifold in the remote
past and returning to $x^{(0)}$ along the contracting manifold in the far future. As suggested by Fig.~\ref{rangeS}, it is
possible to use these large solitonic trajectories to construct periodic orbits, at least for regular tori close
to the singular one.

\subsection{Periodic flows.} 

The Hamiltonian generating the periodic flow  is a function of the $K_l, L_l$: 
$$
{\cal H} = {\cal H}( K_1,L_1, \cdots , K_m, L_m)
$$
This function however is not defined globally and this gives rise to the {\em monodromy} phenomenon.
The flow itself is a linear combinations of the flows generated by $K_l,L_l$
\begin{equation}
\partial_t = \sum_l \alpha_l \partial_{a_l} + \beta_l \partial_{\theta_l}, \quad \alpha_l = {\partial {\cal H}/ \partial K_l}, \quad \beta_l = {\partial {\cal H} / \partial L_l},
\label{periodictime}
\end{equation}
The $X$'s become functions of time
\begin{equation}
X_k = X_k(0) e^{\alpha_k t + i\beta_k t}, \quad X_{\bar{k}} =X_{\bar{k}}(0) e^{-\alpha_k t +i\beta_k t}
\label{Xk}
\end{equation}
We want to find conditions on $\alpha_k, \beta_k$ in order to get a closed trajectory.

To simplify notations,  we will consider the large motion generated by $\alpha_1 K_1+ \beta_1 L_1$.  Fig.~\ref{rangeS} shows that we first have to study the corresponding
symplectic map for an initial condition $(z_j^{in},w_j^{in})$ close to the expanding manifold associated to $K_1$ i.e. such that $w_1^{in}$ is small.
The initial conditions and the time lapse $2T$ 
are chosen in such a way that the image $(z_j^{out},w_j^{out})$ is close to the
contracting manifold, i.e. $z_1^{out}$ is small.  In general, $(z_j^{out},w_j^{out})$ is different from $(z_j^{in},w_j^{in})$
for $j \neq 1$, reflecting the impossibility to express our dynamical system in the vicinity of the pinched torus as a product of $m$ independent
integrable systems with 2 degrees of freedom, each of them exhibiting an isolated focus-focus singularity.
Although such factorization holds at the
topological level~\cite{Zung96}, it does not hold at the level of symplectic manifolds. The obstruction to such a symplectic product decomposition
is encoded in the non-diagonal symplectic invariants, which we will now compute.  For this purpose we will use the fact recalled in Section~\ref{def_symplectic_invariants}, that the  map $(z_j^{in},w_j^{in}) \to (z_j^{out},w_j^{out})$ is strongly constrained by the
condition that $K_1 $ and $L_1$ commute with all the generators of the moment map.

It is first necessary  to identify the components ${\cal C}_{\cal S(\cal I)}^{contracting}$  of the cone  corresponding 
to the $in$ and $out$ states for a large motion generated 
by $\alpha_1 K_1+ \beta_1 L_1$. 
This flow sends the solitonic coordinate $X_1$ into $e^{\alpha_1 a_1+i\beta_1 \theta_1 }X_1$ and $X_{\bar{1}}$ into $e^{-\alpha_1 a_1+i\beta_1 \theta_1 }X_{\bar{1}}$,
and leaves unchanged all the other coordinates $X_k$, $X_{\bar{k}}$. Let us choose an initial condition on an expanding subspace for this flow.
This means that $|X_{\bar{1}}^{{\rm in}}| >> |X_1^{{\rm in}}|$. In principle, the other coordinates can be chosen arbitrarily, but it is convenient to assume also
that  $|X_{\bar{k}}^{{\rm in}}| >> |X_k^{{\rm in}}|$ for $k > 1$, so that the initial point is close to the critical point. Let us denote by ${\cal I}_0$ the set of indices $\{2\cdots m\}$. The manifold ${\cal C}^{in}$ is obtained by sending $a_1$ to $-\infty$ and the manifold ${\cal C}^{out}$  by sending $a_1$ to $+\infty$, the other $j\in \bar{\cal I}_0$ being spectators.
Then ${\cal C}^{in} = {\cal C}^{contracting}_{{\cal S}(\{ \bar{1} \} \cup \bar{\cal I}_0) }$ and ${\cal C}^{out} = {\cal C}^{contracting}_{{\cal S}(\{ 1 \} \cup \bar{\cal I}_0) }$.

The symplectic invariants    are smooth functions
of the conserved quadratic Hamiltonians $K_j,L_j$ ($1 \leq j \leq m$) defined in the vicinity of the singular torus~\cite{Dufour94,San03,Dullin07}. 
The knowledge of solitonic trajectories allows
us to compute them exactly on the singular torus, that is for $K_j=L_j=0$.

\subsubsection{Diagonal case.}
\label{sect_n=n}

As apparent of Fig.[\ref{rangeS}], the trajectory is composed of two pieces. The first one is the large solitonic motion generated by $\alpha_1 K_1 + \beta_1 L_1$
during a time lapse $2T$ which yields a relation between 
$(z_1^{in},w_1^{in})\to (z_1^{out},w_1^{out})$ as explained in Section \ref{def_symplectic_invariants}. This part of the trajectory is  complemented by a further action of the 
flow $\alpha_1 K_1+\beta_1 L_1$ during a time lapse $\tau$ relating 
$(z_{1}^{out}, w_{1}^{out}) \to (z_1^{in}, w_1^{in})$ so as to get a closed trajectory.
This last part of the analysis lies entirely in the domain
of validity of the normal forms. One should therefore have
\begin{equation}
z_1^{in} = e^{\alpha_1 \tau +i \beta_1 \tau } z_1^{out},\quad w_1^{in} = e^{-\alpha_1 \tau +i \beta_1 \tau } w_1^{out}
\label{extraflow}
\end{equation}
Using the leading order expressions for $z_1^{out}, w_1^{out}$, eqs.(\ref{diagonalconditionz},\ref{diagonalconditionw}) 
these two equations are equivalent to
\begin{equation}
\bar{z}_1^{in} w_1^{in} = K_1+iL_1=  e^{-\alpha_1 \tau +i \beta_1 \tau }    \Phi^0_1 (T)
 \label{a1theta1}
\end{equation}

As shown in Section~\ref{def_symplectic_invariants},
one has, again to leading order
$$
w_1^{out} = {\Phi_1^0(T)\over K_1+iL_1} w_1^{in} =  {\Phi_1^0(T)\over \bar{z}^{in}_1} 
$$
The variables  $z_1^{in}, w_1^{in}$ are obtained from $C(E_1),C(E_{\bar{1}})$  
evaluated at $t=-T$ and the variables $z_1^{out}, w_1^{out}$ are obtained from $C(E_1),C(E_{\bar{1}})$ evaluated at $t=T$.  Explicitely we have:
$$
\cases{ C(E_1)\vert_{-T }= \bar{w}_1^{in} \simeq 0  \cr  C(E_{\bar{1}})\vert_{-T } = i a'(\bar{E}_1 ) \bar{z}_1^{in}   } , \quad \cases{ C(E_1)\vert_{  T } = \bar{w}_1^{out}   \cr  C(E_{\bar{1}})\vert_{ T } = i a'(\bar{E}_1 ) \bar{z}_1^{out}  \simeq 0 }
$$
so that:
$$
\Phi_1^0 = w^{out}_1 \bar{z}^{in}_1 = {1\over  i a'(\bar{E}_1 ) } C(E_{\bar{1}})\vert_{-T }  \;   \overline{C(E_1)}\vert_{T }  
$$
When $t\to -T$ the set of $X$'s tending to $\infty$ is ${\cal I} = \overline{\cal I}_0 \cup \{ \bar{1} \}$, then:
$$
C(E_{\bar{1}})\vert_{X_{\bar{1}}\to \infty} \simeq -2 { (E_{\bar{1}} - E_{1})^2\over \prod_{i=1}^n (E_{\bar{1}}-\epsilon_i)} {\prod_{k\in {\cal I}_0}( E_{\bar{1}}-E_k )^2\over X_{\bar{1}}}, \quad C(E_{1}) \simeq 0,
$$
Similarly, when $t\to T$ the set of $X$'s tending to $\infty$ is ${\cal I} = \overline{\cal I}_0 \cup \{ 1 \}$, then:
$$
C(E_1)\vert_{X_1\to \infty} \simeq -2 {\prod_{k \in {\cal I}_0} (E_1 - E_{k})^2\over \prod_{i=1}^n (E_1-\epsilon_i)} {(E_1 - E_{\bar{1}})^2\over X_1}, \quad C(E_{\bar{1}}) \simeq 0,
$$
Hence, we get:
$$
\Phi_1^0 = {4\over i a'(E_{\bar{1}}) }  { (E_{\bar{1}} - E_1)^4 \prod_{k\in {\cal I}_0} (E_{\bar{1}} - E_{\bar{k}})^2 (E_{\bar{1}} - E_k)^2  \over   \prod_{i} (E_{\bar{1}} - \epsilon_i)^2}{1\over X_{\bar{1}}(-T) \overline{X_1}(T)}
$$
Next, we have
$$
 \cases{X_1(T) = X_1(0)  e^{\alpha_1 T +i\beta_1 T} \cr 
 X_{\bar{1}}(-T) = X_{\bar{1}}(0) e^{\alpha_1 T - i\beta_1 T}} \quad \Longrightarrow  \quad X_{\bar{1}}(-T) \overline{X_1(T)} = -{1\over 4} e^{2\alpha_1 T - 2 i \beta_1 T}
$$
so that eq.(\ref{a1theta1}) becomes
$$
c_1\equiv \bar{z}_1 w_1 = e^{-\alpha_1\tau+ i\beta_1 \tau} \Phi_1^0(T) = \rho_1 e^{i \gamma_1} \; e^{-\alpha_1 (2T+\tau) + i\beta_1 (2T+\tau)}
$$
where we have set
\begin{equation}
\rho_1 e^{i\gamma_1} = - {16\over i a'(E_{\bar{1}}) }  { (E_{\bar{1}} - E_1)^4 \prod_{k > 1} (E_{\bar{1}} - E_{\bar{k}})^2 (E_{\bar{1}} - E_k)^2  \over   \prod_{k} (E_{\bar{1}} - \epsilon_k)^2}
\label{eqrho1}
\end{equation}
The quantity $2T+\tau$ is the total time it takes to run once along the closed trajectory, that is the period of the motion. Normalizing it to $2\pi$, the above equation determines $\alpha_1$ and $\beta_1$:
\begin{equation}
\alpha_1 = -{1 \over 2\pi} \log {|c_1| \over \rho_1}, \quad \beta_1 = {1\over 2\pi} ({\rm arg}(c_1)-\gamma_1)
\label{alphabeta1}
\end{equation}

\subsubsection{Non diagonal case.}

Similarly, one can  compensate the non-diagonal motion induced by the large $\alpha_1 K_1+ \beta_1 L_1$ flow by adding  flows $\alpha_j K_j + \beta_j L_j$. The evolution of $z_j$ is now the composition of this new flow
and the indirect evolution due to the $\alpha_1 K_1+ \beta_1 L_1$ flow. Hence
$$
z_j^{out} = e^{(\alpha_j+i\beta_j)(2T+\tau)} \rho_{zj}^{(1)} z_j^{in}, \quad j\neq 1
$$
Periodicity requires 
\begin{equation}
e^{-2\pi (\alpha_j+i\beta_j)} = \rho_{zj}^{(1)}
\label{alphabetaj}
\end{equation}
where we have set the period $2T+\tau$ to $2\pi$. This determines the remaining constants $\alpha_j,\beta_j$ in eq.(\ref{periodictime}), once we have computed 
  $\rho_{zj}^{(1)}$ and $\rho_{wj}^{(1)}$ ($j \neq 1$) as defined in Section~\ref{def_symplectic_invariants}.

The initial conditions were chosen such that when $t=-T$ the system lies  on a subspace of the tangent cone whose natural coordinates  
are the $B(E_j)^{{\rm in}}$'s given by eq.~(\ref{B_expanding}) with ${\cal I} = {\cal I}_0 \cup \{1\}$. In the long time limit $t=T$, i.e. when $a_1= \alpha_1 t$ is positive and large, we get $|X_{\bar{1}}^{{\rm out}}| << |X_1^{{\rm out}}|$,
and the system lies on a different subspace of the tangent cone. The $B(E_j)^{{\rm out}}$ are again given by eq.~(\ref{B_expanding}) with
 ${\cal I} = {\cal I}_0 \cup \{\bar{1}\}$.
From the discussion in Section~\ref{def_symplectic_invariants}, we expect that:
$$
B(E_j)^{{\rm out}}=\rho_{zj}^{(1)} B(E_j)^{{\rm in}}, \quad j\neq 1 
$$
By using eq.(\ref{B_expanding}), this is indeed the case and we find :
\begin{equation}
\rho_{zj}^{(1)}=\left(\frac{E_j-E_1}{E_j-E_{\bar{1}}}\right)^{2}
\label{rhozkj}
\end{equation}
As explained in Section~\ref{def_symplectic_invariants}, we expect the general relation $\rho_{zj}^{(1)}\bar{\rho}_{wj}^{(1)}=1$ between $\rho_{zj}^{(1)}$ and $\rho_{wj}^{(1)}$.
As a consistency check, we may evaluate  $\rho_{wj}^{(1)}$ directly. Using that $w_j=\overline{C(E_j)}$, we have
$$
C(E_j)^{{\rm out}}=\bar{\rho}_{wj}^{(1)} C(E_j)^{{\rm in}}, \quad j\neq 1
$$
Now $C(E_j)^{{\rm in}}$ is obtained by eq.(\ref{C_contracting}) with the index set  ${\cal I}=\overline{{\cal I}_0} \cup \{\bar{1} \} $ and 
 $C(E_j)^{{\rm out}}$ is obtained from the same formula with an index set ${\cal I}=\overline{{\cal I}_0}\cup \{ 1 \}$.
So we have:
\begin{equation}
\bar{\rho}_{wj}^{(1)}=\frac{C(E_j)^{{\rm out}}}{C(E_j)^{{\rm in}}}=\left(\frac{E_j-E_{\bar{1}}}{E_j-E_1}\right)^{2}
\label{rhowkj}
\end{equation}
Clearly, the relation $\rho_{zj}^{(1)}\bar{\rho}_{wj}^{(1)}=1$ is satisfied. Finally the constants $\alpha_j,\beta_j$ are determined by:
\begin{equation}
e^{-2\pi (\alpha_j+i\beta_j)} = \left(\frac{E_j-E_1}{E_j-E_{\bar{1}}}\right)^{2}
\label{alphabetajbis}
\end{equation}

\subsection{ Action variables and symplectic invariants. }
\label{symplectic_invariants}
Let us define  the $1$-form on the base (i.e. the image of the moment map)
$$
\Omega^{(1)} = \sum_j \Omega_j^{(1)}, \quad \Omega_j^{(1)}=\alpha_j dK_{j} + \beta_j dL_{j}
$$
 Remembering that 
$$
c_1 = \bar{z}_1^{in} w_1^{in} = \bar{z}_1^{out} w_1^{out} = K_1+iL_1
$$
we see that $\Omega^{(1)}$ is singular when $c_1\to 0$. Isolating the singular part, we may write
$$
\Omega^{(1)}  =- {1\over 4\pi} ( \log c_1 \; dc_1 + \log \bar{c}_1 \; d\bar{c}_1 ) + \Omega^{(1) reg}
$$
where
$$
\Omega^{(1) reg}= {1\over 2\pi} \log \rho_1\; dK_1 - {1\over 2\pi} \gamma_1\; dL_1 + \sum_{j>1}\alpha_j dK_{j} + \beta_j dL_{j}
$$
The form $\Omega^{reg}$ contains all the regularized symplectic invariants of San V\~{u} Ngoc \cite{San03}, computed here on the singular fiber.
In the case of one spin, we find
$$
\rho_1e^{i\gamma_1} = -{8\over s} (2s-\epsilon_1)^{3/2} (\epsilon_1 + i\sqrt{2s-\epsilon_1^2})
$$
Setting  $s=1$ and $\epsilon_1=0$ yields
$$
\Omega^{(1) reg} = {1\over 2\pi} \left(  5\log 2 \; dK_1 + {\pi\over 2} \;dL_1 \right)
$$
and we recover the invariants computed in \cite{San10} (up to the normalization of the period to $2\pi$).

To define the form $\Omega^{(1)}$, we need to introduce a cut in the $K_1, L_1$ plane. In this cut plane, the form $\Omega^{(1)}$ is closed and remembering eq.(\ref{periodictime}) we have
$$
\Omega^{(1)} = d{\cal H}^{(1)}
$$
The Hamiltonian ${\cal H}^{(1)}$ is the action associated to the closed trajectory we have constructed. It is defined on the cut plane $K_1,L_1$.
\begin{figure}[h!]
\centering
\includegraphics[height= 8cm]{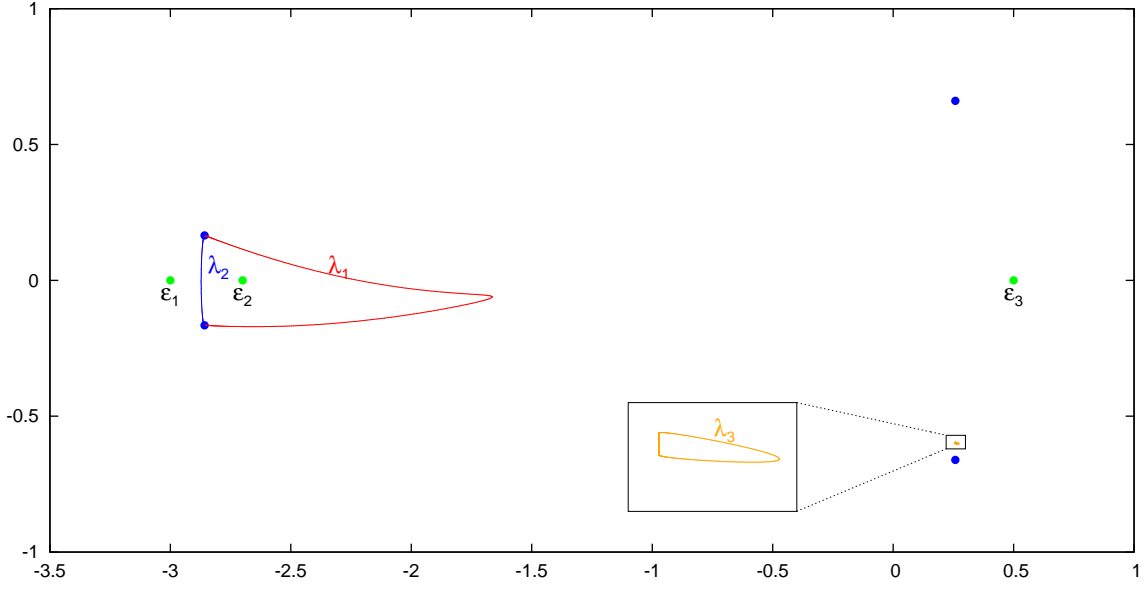}
\caption{The motion of the three separated variables for the soliton solution in the $3$-spins model. The flow is given by eq.(\ref{Xk}) with $\alpha_k,\beta_k$ 
given by eqs.(\ref{alphabeta1},\ref{alphabetajbis}). The limiting flow is obtained by  setting $X_j \to X_j({t\over 2T})$ where $T\simeq -{1\over 4\pi }\log |c_1|$ is large, and $-2\pi T \leq t \leq 2\pi T$.}
\label{diviseur_3spins}
\end{figure}

\begin{figure}[h!]
\centering
\includegraphics[height= 5cm]{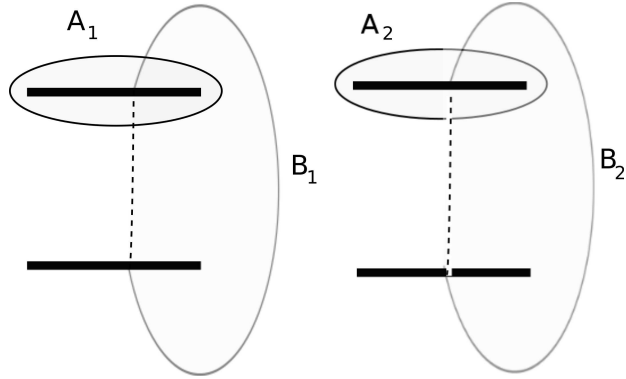}
\caption{The spectral curve with the system of cuts and cycles corresponding to the limiting motion of the separated variables $\lambda_k$.}
\label{coupures}
\end{figure}

We have computed the action ${\cal H}^{(1)}$ to leading order. However following \cite{Dullin07} we can give a more global interpretation using the spectral curve. 
First the normal actions $K_j,L_j$ are just the integrals over the vanishing cycles $A_j$  (encircling pairs of roots of the
spectral polynomial which coalesce into double roots on the critical fiber) as shown in Fig. \ref{coupures}.
\begin{equation}
K_j+iL_j = {1\over 2i\pi} \oint_{A_j} \sqrt{\Lambda(\lambda)} d \lambda
\label{vanishingcycle}
\end{equation}
This can already be seen at the quadratic level
\begin{eqnarray*}
 {1\over 2i\pi} \oint_{A_j} \sqrt{\Lambda} d\lambda &=&  {1\over 2i\pi} \oint_{A_j} \sqrt{ a^2(\lambda) + \sum_k {a(\lambda) \over a'(E_k) (\lambda-E_k)} B_k C_k } \;d\lambda \\
&=& {1\over 2i\pi} \oint_{A_j} \left( a(\lambda)  + {1\over 2}\sum_j {1 \over a'(E_k) (\lambda-E_k)} B_k C_k \right) d\lambda = {1\over 2 a'(E_j)} B_j C_j
= K_j + iL_j
\end{eqnarray*}
Eq.(\ref{vanishingcycle}) has a meaning beyond the quadratic approximation and can be viewed as the inverse of the function $\psi$ in Eliasson normal form \cite{Krich83,Dullin07}.
It is a non linear extension of eqs.(\ref{Kjquad}, \ref{Ljquad}). Concerning the action ${\cal H}^{(1)}$ we see that it has the unique property of being singular 
when $c_1=K_1+i L_1 \to 0$ but it remains regular when $c_j=K_j+i L_j \to 0, j\neq 1$. To get an indication of its global definition, we may consider the example 
of the $3$-spins model and the motion of the separated variables of the soliton solution (see Appendix \ref{sec_n+1_degenerate})  corresponding to the limiting motion generated by ${\cal H}^{(1)}$ on the singular fiber.
On Fig.[\ref{diviseur_3spins}] we show the motion of the separated coordinates $\lambda_k$  defined in eq.(\ref{Csepare}). Here $\lambda_1$ is associated 
to the sign $(+)$ in eq.(\ref{mui}), while $\lambda_2, \lambda_3$ are associated to the sign $(-)$. Hence the corresponding points $(\mu_k,\lambda_k)$  belong to different sheets in the representation of the spectral curve as a covering of the complex plane.  On a fiber close to the singular one, the action integral is
$$
{\cal H}^{(1)} = {1\over 2i\pi} \sum_k \oint \mu_k d\lambda_k ={1\over 2i\pi} \sum_k \oint \sqrt{\Lambda(\lambda_k)} d\lambda_k 
$$
where the $\lambda_k$ describe trajectories close to the critical ones.
When $k=3$ we integrate over a trivial cycle and the contribution vanishes. The $k=1,2$ contributions combine to make a non trivial cycle $B_1$ as indicated on Fig.[\ref{coupures}].
\begin{equation}
{\cal H}^{(1)} = {1\over 2i\pi}  \oint_{B_1} \sqrt{\Lambda(\lambda)} d\lambda 
\label{b_cycle}
\end{equation}
We can eliminate the $H_i$ between eq.(\ref{vanishingcycle}) and eq.(\ref{b_cycle}) thereby obtaining the full Taylor expansion of ${\cal H}^{(1)}$ in terms of the $K_j, L_j$.  

 We believe that this can be generalized to the $n$ spin model.  We have $2m$ vanishing cycles $A_j$ and $2m$ cycles $B_j$. 
The action integrals on the cycles $A_j$ 
give directly the $c_j=K_j+i L_j$ and the action integral on the $B_j$ give the periodic Hamiltonian ${\cal H}^{(j)}$ (see section  \ref{appendix_periodic}) 
which become singular when $c_j$ tends to zero but remain regular when $c_i, i \neq j$ tends to zero.
The important ingredient here is the identification of the $B_j$ cycles through the motion of the divisor of the proper soliton solution. 
 A complete analysis of this motion has not yet been achieved, however using the asymptotic expressions for determinants given in section~\ref{asymptotic},
it is possible to show that as the time $t_1$ goes to $-\infty$, we have one separated variable $\lambda_{1}^{+}$ located at $E_{\bar{1}}$, and another one, $\lambda_{1}^{-}$, at
$E_1$. When  $t_1$ goes to $+\infty$, there is  one separated variable $\lambda_{i}^{+}$ located at $E_{1}$ and one $\lambda_{j}^{+}$ at $E_{\bar{1}}$, 
The other remaining separated variables, $m-2$ of type $\lambda^{+}$
and $m-1$ of type $\lambda^{-}$ are found to occupy the same sets of points in the complex plane in both limits $t_1 \rightarrow -\infty$ and  $t_1 \rightarrow \infty$ due to the periodicity conditions eq.(\ref{alphabetajbis}).
The difficult task remains to show that this possibly complicated motion of the separated variables can be continuously deformed into one of the $B_j$ cycles.
If one succeeds in doing this, the periodic Hamiltonians can then be obtained more globally by the construction of section \ref{appendix_periodic}. 
An alternative approach would be to use Picard-Fuchs equations for the action variables as in \cite{PaDu12}.

\subsection{Monodromy} 

The orbits of the periodic motions defined in the previous section provide a basis of cycles on the Liouville tori close to the singular fiber. It is well known that 
when we follow these cycles by continuity when we perform a non trivial closed contour in the image of regular values of the moment map, the initial and final basis of cycles 
may be different and are related by an element of  $GL(n,\mathbb{Z})$.

In order to compute the monodromy, we have first to determine the fundamental group $\pi_1$ of the manifold of regular values of the moment map.
The manifold of singular values of the moment map is a stratified manifold by the rank of the moment map.  At the equilibrium point the rank  is zero, however 
the equilibrium point belongs to bigger strata where the rank is $r \leq 2m$. Eliasson's theorem allows an easy  description of this  stratification in a local neigborhood $U$ of the equilibrium point. The moment map is given by 
$$
(x,p) \in U \to (K_1,L_1,K_2,L_2,\cdots , K_m,L_m)
$$
The derivatives of each pair $K_j,L_j$ generically span a two dimensional subspace of the tangent space of the image of the moment map near the equilibrium point  and these spaces are linearly independent for two different pairs. 
For each pair the only way  to span a space of dimension smaller than $2$ is when the two tangent vectors  are identically zero. In this case the rank of the momentum map drops to 
$r=2m-2$ and the equation of this stratum is $K_j=L_j=0$. The singular values of the momentum map in a neighborhood of the equilibrium point  consists  therefore 
of $m$ strata ${\cal O}_j$ 
of dimension $2m-2$ given by the equations $K_j=L_j=0$, for  $i\in \{1,\cdots m\}$. Of course each one of these strata is itself stratified by the intersection of $p$ such 
manifolds ${\cal O}_{j_1} \cap {\cal O}_{j_2}\cap \cdots {\cal O}_{j_p}  $ where the rank is $r=2m-2p$. The equilibrium point itself is obtained when $p=m$.

Because the co-dimension of  ${\cal O}_{j}$ is $2$, we can draw non trivial circles $S_1$ around it. Indeed ${\cal O}_{j}$ intersects the plane $K_j,L_j$ precisely at the origin and such an   $S_1$  can be taken to be $K_j+ i L_j \to e^{i\alpha_j} (K_j+ i L_j)$.
The fundamental group of the image of the momentum map minus the ${\cal O}_{j}$'s is a product of $m$ factors $S_1$: $\pi_1= S_1 \times S_1 \times \cdots S_1$.

We can now  compute the monodromy. 
Locally we have $\Omega^{(1)} =d {\cal H}^{(1)} $.
Recall that 
$$
\bar{z}_1 w_1 = K_1+i L_1 = \rho_1 e^{i\gamma_1}
$$
The regular values of the moment map are such that $\rho_1 \neq 0$. Monodromy is defined  in the open set of the regular values of the moment map. 
Let us take as a non trivial closed path  a circle around the origin in the $K_1,L_1$ plane: $c_1(\alpha_1)= e^{i\alpha_1}\rho_1$, $\gamma_1\leq \alpha_1 \leq \gamma_1+2\pi$.
Integrating the form $\Omega^{(1)}$   over this circle one    gets a non trivial contribution from $\Omega_1^{(1)}$
\begin{eqnarray*}
\oint_C \Omega^{(1)} =\oint_C \Omega_1^{(1)}&=&-{ i\over 4\pi} \oint_{\gamma_1}^{2\pi + \gamma_1} \Big[c_1(\alpha_1)\log c_1(\alpha_1)- \bar{c}_1(\alpha_1) \log \bar{c}_1(\alpha_1) \Big] d\alpha_1 \\
&=&
 {1\over 2\pi} \rho_1 \oint_{\gamma_1}^{2\pi + \gamma_1} \alpha_1 \cos \alpha_1  d\alpha_1
= \rho_1 \sin \gamma_1 = L_1
\end{eqnarray*}
This means that when we go around that circle, the Hamiltonian ${\cal H}^{(1)}$ becomes ${\cal H}^{(1)} + L_1$.  The new periodic trajectory contains 
an extra  $S^1$ factor generated by $L_1$. This is exactly the monodromy phenomenon. Note that when the image of the moment map winds once around the stratum $\mathcal{O}_j$, $\mathcal{H}^{(1)}$ remains single-valued when $j \neq 1$, so for $\mathcal{H}^{(1)}$ only paths which wind around the stratum $\mathcal{O}_1$ give rise to a non-trivial monodromy.

Following up the transformation of the action integrals ${\cal H}^{(j)}$ defined in the previous section along such a non trivial path  would also lead to an interpretation of the monodromy in the framework of the Picard-Lefschetz theory as in \cite{Audin01}.

\section{Appendix : Normal Forms.}
\label{sec_Normal_Forms}
Normal coordinates  simultaneously diagonalize all the flows $H_j$ in the quadratic approximation.
\begin{equation}
\{ H_j, a_\alpha \} = E_{j,\alpha} a_\alpha
\label{eqnormal}
\end{equation}
To construct them, we return to  eqs.(\ref{AB} -- \ref{BC}). They imply
$$
\left\{ {1\over 2}{\rm Tr}\; L^2(\lambda), C(\mu) \right\} = {2i\over \lambda-\mu} \Big( A(\lambda) C(\mu) - A(\mu) C(\lambda) \Big)
$$

When we expand around a configuration given by eq.(\ref{static}), all the quantities ($b$, $\bar{b}$, $s_j^+$, $s_j^-$) 
are first order, but $s_j^z$ is second order because $ s_j^z = s e_j -{e_j\over 2s} s_j^+ s_j^-  + \cdots, \quad e_j = \pm 1$. 
In this expansion, the Hamiltonians are quadratic while $C(\mu)$ is first order.  Therefore the Poisson bracket in the  left-hand side is linear. Now $A(\lambda)$ is constant plus second order, so that in the right-hand side we can replace $A(\lambda)$ and $A(\mu)$ by their 
zeroth order expression :
$$
A(\lambda) \simeq a(\lambda) = 2\lambda + \sum_{j=1}^{n} {s e_j \over \lambda-\epsilon_j}
$$
and  we arrive at
\begin{equation}
\left\{ {1\over 2}{\rm Tr}\; L^2(\lambda), C(\mu) \right\} = {2i\over \lambda-\mu} \Big( a(\lambda) C(\mu) - a(\mu) C(\lambda) \Big)
\label{eq1}
\end{equation}
Eq.(\ref{eq1})  will be precisely of the form of eq.(\ref{eqnormal}) if we can kill the unwanted term $C(\lambda)$. This is achieved by imposing the condition 
\begin{equation}
a(\mu)=0, \quad  \mbox{ ``Classical Bethe Equation''}
\label{classicalBethe}
\end{equation}
This is an equation of degree $n+1$ for $\mu$. For non degenerate singularities it has only simple roots. Calling $ E_i$ its solutions,  we construct in this way $n+1$ variables $C( E_i)$. Clearly,  they all Poisson commute
\begin{equation}
\{ C( E_i), C( E_j) \} = 0
\label{CiCj}
\end{equation}

 Since phase space has dimension $2(n+1)$ this is half what we need. To construct the conjugate variables, we consider eq.(\ref{BC}). In our linear approximation it reads
$$
\{ B( E_i), C( E_j)\} = {2i\over  E_i- E_j} (a( E_i)-a( E_j) )
$$
If $ E_i$ and $ E_j$ are {\it different } solutions of eq.(\ref{classicalBethe}), then obviously
\begin{equation}
\{ B( E_i), C( E_j)\} =0, \quad  E_i \neq  E_j
\label{BiCj}
\end{equation}
If however $ E_j= E_i$ then
\begin{equation}
\{ B( E_i), C( E_i)\} =  2i a'( E_i)
\label{BiCi}
\end{equation}
Finally,  we also have
\begin{equation}
\{ B( E_i), B( E_j)\} = 0
\label{BiBj}
\end{equation}
Up to normalisation, we have indeed constructed canonical coordinates.

\bigskip

It is simple to express the quadratic Hamiltonians in theses coordinates:

\begin{equation}
{1\over 2}{\rm Tr}\; L^2(\lambda) = a^2(\lambda) + \sum_j { a(\lambda)\over a'( E_j) (\lambda- E_j) } B( E_j) C( E_j)
\label{benoit}
\end{equation}
This has the correct analytical properties in $\lambda$ and together with  the Poisson brackets eqs.(\ref{CiCj},\ref{BiCj}, \ref{BiCi}, \ref{BiBj})  we reproduce eq.(\ref{eq1}). Note that there is no pole at $\lambda= E_j$ because $a( E_j)=0$.  Expanding around $\lambda=\infty$ we get
$$
H_{n+1}= s\sum_k e_k + \sum_i {1\over 2 a'( E_i)} B( E_i) C( E_i)
$$
and computing the residue at $\lambda=\epsilon_j$, we find
$$
H_j= se_j\left[ 2\epsilon_j + \sum_k {se_k \over \epsilon_j-\epsilon_k}  \right] + \sum_i {1\over 2 a'( E_i)} { se_j \over \epsilon_j- E_i} B( E_i) C( E_i) 
$$
We can invert these formulae: devide eq.(\ref{benoit}) by $\lambda- E_j$ and take the residue at $\lambda= E_j$. Since $a( E_j)=0$ we get
$$
{1\over 2}{\rm Tr}\; L^2( E_j) =  B( E_j) C( E_j)
$$
or explicitly
$$
 B( E_j) C( E_j) = 4 E_j^2 + 4 H_{n+1} + \sum_{i=1}^n {2H_i\over  E_j-\epsilon_i} + \sum_{k=1}^n {s^2\over ( E_j-\epsilon_i)^2}
 $$

\noindent
We can now make contact with the Williamson classification theorem.

\bigskip

If $ E_j$ is {\em real} we have $B( E_j)=\overline{C( E_j)}$ and we set
$$
C( E_j) = \sqrt{ |a'( E_j) |} (p_j+i \eta_j  q_j), \quad B( E_j) = \sqrt{ |a'( E_j) |} (p_j-i\eta_j q_j )
$$
where $\eta_j = - {\rm sign}(a'( E_j))$ and $p_j, q_j$ are {\em real} canonical coordinates. Then
eq.(\ref{BiCi}) is satisfied. Moreover 
$$
B( E_j) C( E_j) =  \vert a'( E_j) \vert (p_j^2+ q_j^2 ) 
$$
 i.e. we have an {\em elliptic} singularity.

\bigskip

If $ E_j$ is {\em complex}, there is another root $ E_{j+1}= \bar{E}_j$ which is its complex conjugate.
Then $\overline{B( E_j)}=C( E_{j+1})$. We introduce {\em real} canonical coordinates $p_j,q_j,p_{j+1},q_{j+1}$ and  set
$$
B( E_j) = -ia'( E_j) (q_{j} + i q_{j+1}) ,\quad B( E_{j+1} )= p_j + i p_{j+1}
$$
$$
C( E_j) = p_j -i p_{j+1},\quad C( E_{j+1})= i a'( E_{j+1})  (q_j -i q_{j+1} )
$$
then eqs.(\ref{CiCj},\ref{BiCj},\ref{BiCi},\ref{BiBj}) are satisfied and
\begin{eqnarray*}
B( E_j)C( E_j) &=& -i a'( E_j) (K_j  +i L_j ) \\
B( E_{j+1})C( E_{j+1}) &=& i a'( E_{j+1}) (K_j  - i L_j ) 
\end{eqnarray*}
i.e. we have a {\em focus-focus} singularity.

\bigskip

\section{Appendix: In-Out transformation.}
\label{def_symplectic_invariants}
We consider a flow 
generated by the Hamiltonian $\alpha_1 K_1 + \beta_1 L_1$. Such linear combinations were  needed in Section \ref{comp_symplectic_invariants}, although here only the $K_1$ part really matters. We assume $\alpha_1 >0$.
Let us fix some  $\delta >0$ with $\delta$ small enough that Eliasson map $\Phi$ is well defined in the ball of radius $2\delta$ around
the origin. Consider the point  $U:  (z_1 = \delta, w_1=0 )$ on the expanding manifold and the trajectory of the
flow starting at that point. 
Since the trajectory stays on the {\em global} unstable manifold, which is equal to the {\em global} stable manifold, it will reappear 
after a
time lapse 
$2T$ 
on the contracting manifold  at a point $S : ( z_1= 0,  |w_1| = \delta) $. The
flow evaluated at  time $2T$ is well defined in 
a neighborhood $\Sigma_U$ of the point $U$ and defines a symplectic mapping $F^{(1)}_T$ 
of   $\Sigma_U$,  $|w_{1}^{{\rm in}}| << |z_{1}^{{\rm in}}|$,  
to a neighborhood $\Sigma_S$ of the point $S$ in the contracting subspace $|z_{1}^{{\rm out}}| << |w_{1}^{{\rm out}}|$:
$$
\Sigma_U:(z_1^{in},w_1^{in}, z_2^{in},w_2^{in}, \cdots , z_m^{in},w_m^{in})  \stackrel{F^{(1)}_T}{\longrightarrow} \Sigma_S : (z_1^{out},w_1^{out},z_2^{out},w_2^{out}, \cdots , z_m^{out},w_m^{out}) 
$$
We also assume  that the other components  $z_{i}^{{\rm in}}$,
$w_{i}^{{\rm in}}$,  $z_{i}^{{\rm out}}$, $w_{i}^{{\rm out}}$, $i=2,\cdots, m$ are small (this will be satisfied in our case) 
so that the initial and final points lie in a small neighborhood of the critical point.
One can find the precise form of this mapping by writing that it commutes with the flows generated by $K_j,L_j$, $j= 1\cdots m$.
Let us fix the level set by imposing: 
\begin{equation}
\bar{z}_j^{in} w_j^{in}= \bar{z}_j^{out} w_j^{out} = K_j+i L_j
\label{in=out}
\end{equation}
Let us also assume that $ K_j+i L_j \neq 0$. We may then choose $(w_1^{in},\cdots,w_m^{in})$ as coordinates on the intersection of $\Sigma_U$
with the level set and write:
\begin{eqnarray*}
w_j^{out} = F_j^{(1)}(K_1,L_1,\cdots K_m,L_m,w_1^{in},\bar{w}_1^{in},\cdots, w_m^{in},\bar{w}_m^{in} ) 
\end{eqnarray*}
The condition that the transformation $F^{(1)}$ commutes with the flows $K_1,L_1,\cdots,K_m,L_m$  means:
\begin{eqnarray*}
e^{-a_j + i\theta_j}  F_j^{(1)}(K_1,L_1,\cdots K_m,L_m,w_1,\bar{w}_1,\cdots, w_m,\bar{w}_m ) &=& \\
 &&  \hskip -6cm F_j^{(1)}(K_1,L_1,\cdots K_m,L_m,e^{-a_1 + i\theta_1}  w_1, e^{-a_1 - i\theta_1}  \bar{w}_1,\cdots, e^{-a_m + i\theta_m}  w_m,e^{-a_m - i\theta_m}  \bar{w}_m ) 
\end{eqnarray*}
To solve this equation we choose the parameters $a_k,\theta_k$ such that
$$
e^{-a_k + i\theta_k}  w_k = \eta_k, \quad  w_k, \eta_k \neq 0, \quad 1\leq k \leq m
$$
where $\eta_k$ are constants. Then the equation becomes:
\begin{eqnarray*}
 F_j^{(1)}(K_1,L_1,\cdots K_m,L_m,w_1,\bar{w}_1,\cdots, w_m,\bar{w}_m ) = \rho_{wj}^{(1)}(K_1,L_1,\cdots K_m,L_m) w_j
\end{eqnarray*}
where
$
\rho_{wj}^{(1)}(K_1,L_1,\cdots K_m,L_m ) =  \eta_j^{-1}F_j^{(1)}(K_1,L_1,\cdots K_m,L_m,\eta_1,  \bar{\eta}_1,\cdots, \eta_m, \bar{\eta}_m )
$.
Hence, we have shown that
\begin{equation}
w_j^{out} =  \rho_{wj}^{(1)}(K_1,L_1,\cdots K_m,L_m) w_j^{in}
\label{defrhow}
\end{equation}
where the super script $(1)$ is to remind that this is a transformation under the $\alpha_1 K_1 +\beta_1 L_1$ flow. In exactly the same way, we show
\begin{equation}
z_j^{out} =  \rho_{zj}^{(1)}(K_1,L_1,\cdots K_m,L_m) z_j^{in}
\label{defrhoz}
\end{equation}
The conditions eq.(\ref{in=out}) then imply
\begin{equation}
\rho_{wj}^{(1)}(K_1,L_1,\cdots K_m,L_m )  \overline{\rho_{zj}^{(1)}(K_1,L_1,\cdots K_m,L_m) } =1
\label{consistency_rho}
\end{equation}
When $j \neq 1$, the functions $\rho_{wj}^{(1)}(K_1,L_1,\cdots K_m,L_m )$ do not depend on the choice of the time lapse $2T$
for the
flow, nor on the choice of the parameters $\alpha_1,\beta_1$. They are smooth in a neighborhood of the critical torus.
As shown before~\cite{San03,Dullin07} they are symplectic invariants attached to the Liouville foliation in the vicinity of the
singular torus. Let us now show that for $j=1$, $\rho_{w1}^{(1)}(K_1,L_1,\cdots K_m,L_m )$ is singular as $K_1+iL_1 \to 0$.
 
To see this, we first consider the critical torus, on which the dilating stratum is given by $w_1=0$ and the contracting stratum by $z_1=0$.
We start with an initial condition on  the dilating stratum $(w_{1,sol}^{in}= 0, z_{1,sol}^{in})$. 
We let the system evolve with respect to 
the flow
. After a large motion during the time lapse $2T$, 
the point reappears on the contracting stratum $(w_{1,sol}^{out}, z_{1,sol}^{out} = 0)$.
The same reasoning as before on the invariance under the flows $K_j,L_j$ shows that:
\begin{equation}
w_{1,sol}^{out} = {\Phi_1^0 \over \bar{z}_{1,sol}^{in}}
\label{inout}
\end{equation}
where $\Phi_1^0$ is a constant, which depends only on the combination $\alpha_1 T$. 
Let us now consider a torus close to the critical torus. 
The point $(z_{1,sol}^{in}, w_{1,sol}^{in}=0)$ is displaced to $(z_1^{in}, w_1^{in})$ and  $(z_{1,sol}^{out}=0, w_{1,sol}^{out})$ is displaced 
to $(z_1^{out}, w_1^{out})$. The small parameters can be chosen to be $w_j^{in}$ ($1 \leq j \leq m$) and we can write:
\begin{eqnarray}
z_1^{out} &=& \left( {  z_1^{in} \bar{w}_1^{in} \over \overline{\Phi^0_1} }\right)  z_1^{in} + O(w^2)
\label{diagonalconditionz} \\
w_1^{out} &=& { \Phi^0_1\over \bar{z}_1^{in}} +  O(1)  w_1^{in} + O(w^2) = \left( {\Phi^0_1\over \bar{z}_1^{in} w_1^{in}} +  O(1)  \right) w_1^{in} + O(w^2) 
\label{diagonalconditionw} 
\end{eqnarray}
where we have imposed the condition $\bar{z}_1^{in} w_1^{in} = \bar{z}_1^{out} w_1^{out} = K_1+iL_1$.
Connecting these equations with the definition~(\ref{defrhow}) we see that:
$$ 
\rho_{w1}^{(1)} \simeq {\Phi_1^0 \over K_1 + i L_1} +  O(1)
$$
in the vicinity of the singular torus. As explained in~\cite{San03}, a regularization procedure is needed in order to define
the diagonal symplectic invariant.

\section{Appendix: Symplectic invariance. }
In this Appendix we explain why the quantity $\Omega^{reg}$ computed in Section \ref{symplectic_invariants} has an intrinsic meaning and is invariant under symplectic transformations preserving the singular Liouville foliation. We extend to higher dimensions the arguments developed in \cite{San03}.

Recall that two singular foliations ${\cal F}$ and ${\cal F}'$ in the symplectic manifolds 
$M,\omega$ and $M',\omega'$ are  equivalent if there exists a symplectic diffeomorphism $\varphi :  {\cal F} \to {\cal F}'$ sending leaves to leaves.

Having two such foliations, by Eliasson theorem, we can define for each of them normal coordinates $(z_j,w_j)$ and $(z_j',w_j')$, $j=1\cdots m$, valid in neighborhoods  $U$ and $U'$ of the focus-focus point.  We can also define the moment maps $\mu^{(0)} : (z_j,w_j) \to (K_j, L_j)$ and  $\mu^{(0)'} : (z_j',w_j') \to (K_j', L_j')$, $j=1\cdots m$ where 
 the Hamiltonians $K_i, L_i$ and  $K_i', L_i'$ are as in Section \ref{section_normal_flows}. In the neighborhoods  $U$ and $U'$, the study of the  symplectic diffeomorphism $\varphi$
 therefore reduces to the study of the commutative diagram
\begin{equation}
\begin{array}[c]{ccc} 
 \mathbb{R}^{4m} &\stackrel{\Phi}{\longrightarrow}& \mathbb{R}^{4m} \\
 \downarrow\scriptstyle{\mu^{(0)}}& &\downarrow\scriptstyle{\mu^{(0)'}}\\ 
 \mathbb{R}^{2m}  &\stackrel{\Psi}{\longrightarrow}& \mathbb{R}^{2m}   
\end{array}
\label{exact1}
\end{equation}
where $\Phi$ is a symplectomorphism and $\Psi $ is a diffeomorphism. We will show that the diffeomorphism $\Psi$ is extremely constrained and such that  
$L_i= \zeta_i L_i'$ with $\zeta_i=\pm 1$ and 
$K_i = K_i'$,  up to a flat function (vanishing at the origin together with all its derivatives) so that the Taylor expansion of $\Omega^{reg}$ has an invariant meaning.

The first observation is that the rank of the moment map is preserved under $\Phi$. Manifolds where the rank of $\mu^{(0)}$ is $r$ are mapped to manifolds where the rank of $ \mu^{(0)'}$ is $r$.  Manifolds of rank $2$ are the (complex) coordinate planes $z_i,w_i$, characterized by the equations $c_i\equiv K_i+i L_i \neq 0$ and $c_j=0$ for  $ j\neq i$. Therefore the map $\Psi$ sends the coordinate axis (in complex notation) $c_i \neq 0$ to a coordinate axis $c'_{i'} \neq 0$. This defines a map $\sigma: i'=\sigma(i)$ of the set $\{ 1,2 \cdots m\}$ to itself which is a permutation. We can assume that this permutation is the identity.

One can generalize this argument. Manifolds of rank $4$ are  the characterized by $c_i \neq 0, c_j \neq 0$ and all other $c_k=0$. Therefore the   map  $\Psi$ sends the plane $c_i \neq 0, c_j \neq 0$  and all other $c_k=0$ to a similar plane  $c_r' \neq 0, c_s'\neq 0$ and all other $c_l'=0$. But that plane
must contain the axes  $c_i' \neq 0$ and $c_j'\neq 0$, hence it is the whole plane  $c_i' \neq 0, c_j' \neq 0$ and all other $c_k'=0$.
Repeating the argument,  manifolds of rank $2m-2$ are hyperplanes $z_i=0, w_i=0$ characterized by the equations $c_i= K_i+i L_i = 0$ and $c_j\neq 0$ for  $ j\neq i$ and the map $\Psi$  sends the hyperplane $c_i=0$ to the hyperplane $c'_i=0$.

The symplectomorphism $\Phi$ sends a periodic flow to a periodic flow of the same period. In particular, the flow generated by $L_i$ is periodic. 
Let us denote by $(K'_i(K,L),L'_i(K,L))$ the coordinates of the image by $\Psi$ of the point of coordinates $(K_i,L_i)$.
The Hamiltonian vector field of $L_i'$ is
$$
X_{L'_i} = \sum _k {\partial L'_i \over \partial L_k } X_{L_k} + {\partial L'_i \over \partial K_k } X_{K_k} 
$$
However, {\em on the open sets $U$ and $U'$}, the only periodic flows are those generated by the $L_k$ and $L_k'$. Hence ${\partial L'_i \over \partial K_k }  = 0$ and $ {\partial L'_i \over \partial L_k } = A_{ik}$ where $A_{ik}$ is an invertible matrix of $GL_m(\mathbb{Z})$. Therefore
$$
L'_i = \sum_k A_{ik} L_k 
$$
On the other hand, we have just seen that if only $c_i$ is different from zero, then only $c'_i$ is different from zero. This implies that the matrix $A$ is diagonal, and since it is an element of $GL_m(\mathbb{Z})$, it is of the form $A_{ik} = \zeta_i \delta_{ij}, \zeta_i=\pm1$. We have shown that
\begin{equation}
L'_i = \zeta_i L_i, \quad \zeta_i = \pm 1
\label{Lprimei}
\end{equation}
Next we want to analyse the functions
$$
K'_i = K'_i( K_1, L_1 \cdots K_m, L_m)
$$
The remarks on the preservation of the rank of the moment map put constraints on the form of the Taylor expansions of these functions. First, because the manifold
$c_i = 0, c_j \neq 0, j\neq i$ is mapped to the manifold $c'_i=0, c_j' \neq 0, j\neq i$, we can write
\begin{equation}
K'_i = A_i( K_1, L_1 \cdots K_m, L_m) K_i +  B_i( K_1, L_1 \cdots K_m, L_m) L_i
\label{Kprimei}
\end{equation}
where the functions $A_i, B_i$ are regular at $0$. Moreover, since the axis $c_i\neq 0, c_j = 0, j\neq i$ is mapped on the axis $c'_i \neq 0, c_j' = 0, j\neq i$ we have
$$
A_i( K_1, L_1 \cdots K_m, L_m) = \alpha_i +O(K,L),\quad B_i( K_1, L_1 \cdots K_m, L_m) = \beta_i  +O(K,L)
$$
where $\alpha_i, \beta_i$ are constants such that $\alpha_i \beta_i \neq 0$.
We show below  the much stronger results $\alpha_i +O(K,L) = 1 + {\rm flat} $ and   $ \beta_i+O(K,L)= 0+  {\rm flat} $.

Let us assume for simplicity that $i=1$. 
Let us consider, in the plane $(z_1',w_1')$, the points on the critical torus ${m'}^{in}_0 : (w'_1=\delta, z_1'=0)$ and ${m'}^{out}_0 : (w'_1=0, z_1'=\delta)$ (all other coordinates vanish). 
The points ${m}^{in}_0= \Phi^{-1}({m'}^{in}_0)$ and ${m}^{out}_0= \Phi^{-1}({m'}^{out}_0)$ are in the plane $(z_1,w_1)$ on the critical torus and we can assume  ${m}^{in}_0 : (w_1=a, z_1=0)$,
${m}^{out}_0 : (w_1=0, z_1=b)$ (all other coordinate vanish). 

Let us now consider a small neighborhood of ${m'}^{in}_0$ and let us apply at each of its points  the flow generated by $(K_1' , L_1')$ during times $(a'_1,\theta'_1)$ chosen in such a way that the point ${m'}^{in} : ({z_1'}^{in}=\bar{c}'_1/\delta, {w'_1}^{in}=\delta, {z_j'}^{in}, {w_j'}^{in})$ is sent to ${m'}^{out} : ({z'_1}^{out}=\delta, {w'_1}^{out}=c'_1/\delta,  {z_j'}^{out}= {z_j'}^{in},  {w_j'}^{out}= {w_j'}^{in})$ i.e. 
$$
e^{-a_1'+i\theta_1'}= {c'_1\over \delta^2}
$$
In particular, when $|c'_1|\to 0$, we find
$$
a_1'\to -\log |c'_1| \simeq -\log |c_1|
$$
where the last relation comes from eqs.(\ref{Lprimei},\ref{Kprimei}).
This map sends ${m'}^{in}_0$ to ${m'}^{out}_0$ and  extends to a smooth map $F'$ from a neighborhood of ${m'}^{in}_0$ to a neighborhood of ${m'}^{out}_0$. The map $F=\Phi^{-1} \circ F' \circ \Phi$ sends a neighborhood of ${m}^{in}_0$ to a neighborhood of ${m}^{out}_0$ and  this map, being a composition of $C^\infty$ maps, should be $C^\infty$.
On the other hand,  the map $F$ is obtained by applying the flow
$$
 \sum_{j=1}^m a'_1 {\partial K'_1 \over \partial K_j} X_{K_j} + \theta'_1 X_{L_1} +  \sum_{j=1}^m a'_1 {\partial K'_1 \over \partial L_j} X_{L_j} 
$$
On $(z_1,w_1)$, this gives
\begin{eqnarray*}
z_1^{out}& =& \exp\left[ a'_1 {\partial K'_1 \over \partial K_1} + i\left( \theta'_1+ a'_1  {\partial K'_1 \over \partial L_1} \right)\right] z_1^{in}\\
w_1^{out}& = & \exp\left[ -a'_1 {\partial K'_1 \over \partial K_1} + i\left( \theta'_1+ a'_1  {\partial K'_1 \over \partial L_1} \right)\right] w_1^{in}
\end{eqnarray*}
When $|c_1|\to 0$, we should have $z^{in} \to {\bar{c}_1\over \bar{a}}$, $z^{out} \to b$. Inserting into the first equation,we find
$$
\bar{a}b = \exp \left[ \log \bar{c}_1 + a'_1 {\partial K'_1 \over \partial K_1} + i\left( \theta'_1+ a'_1  {\partial K'_1 \over \partial L_1} \right)\right]
$$
(for the other equation $w^{in} \to a $ $ w^{out}\to {c_1\over \bar{b}} $, we obtain the same condition). Using the values of $a'_1,\theta'_1$, we get in the limit 
$|c_1|\to 0$
$$
\bar{a}b = \exp \left[ \log |c_1] \left(1- {\partial K'_1 \over \partial K_1} - i  {\partial K'_1 \over \partial L_1} \right)\right]
$$
Finiteness when $ |c_1] \to 0$ requires $\alpha_1=1, \beta_1=0$. One can push the argument and show that
smoothness of the map  in the limit $c_1\to 0$ 
gives
$$
 {\partial K'_1 \over \partial K_1} =1 + {\rm flat}  ,\quad  {\partial K'_1 \over \partial L_1}= {\rm flat}
$$
the argument is the same as in \cite{San03}.

On $(z_j,w_j), j \neq 1$, we get
\begin{eqnarray*}
z_j^{out} &=& \exp\left[ a'_1 {\partial K'_1 \over \partial K_j} + i\left(  a'_1  {\partial K'_1 \over \partial L_j} \right)\right] z_j^{in} \\
w_j^{out} &=& \exp\left[ -a'_1 {\partial K'_1 \over \partial K_j} + i\left(  a'_1  {\partial K'_1 \over \partial L_j} \right)\right] w_j^{in}
\end{eqnarray*}
We require that $(z_j^{out} , w_j^{out})$ be finite when $c_1$ tends to zero with $(z_j^{in} , w_j^{in})$ finite. This gives
$$
 {\partial K'_1 \over \partial K_j} = {\rm flat}  ,\quad  {\partial K'_1 \over \partial L_j}= {\rm flat}
$$

\section{Appendix: Solitons.}
\label{sec_n+1_degenerate}
In this appendix  we recall the construction of solitons in the Jaynes-Cummings-Gaudin model~\cite{Yuzbashyan08}. We present it in some details because we need to generalize it to include all times $t_i$ and we wish to present in a self contained way the formulae  we use in our calculation of the symplectic invariants.

The Lax form eq.~(\ref{eqLax}) of the equation of motion implies that the so-called
spectral curve $\Gamma$, defined by $\det ( L(\lambda) - \mu) = 0$ is a constant of motion.  
Specifically:
\begin{equation}
\Gamma: \;   \mu^2 - A^2(\lambda) - B(\lambda) C(\lambda) = 0, \textrm{ i.e. } \mu^2 =  {Q_{2n+2}(\lambda) \over \prod_j (\lambda - \epsilon_j)^2}
\label{QABC}
\end{equation}
Defining $y = \mu  \prod_j (\lambda - \epsilon_j)$,  the equation of the curve becomes $y^2 = Q_{2n+2}(\lambda)$
which is an hyperelliptic curve. Since the polynomial  $Q_{2n+2}(\lambda)$ has degree $2n+2$,  the genus of the curve in $n$. 

The separated variables are $g=n$ points  on the curve whose coordinates $(\lambda_k,\mu_k)$ can be taken as coordinates on phase space. 
They are defined as follows. Let us write
\begin{equation}
C(\lambda) = 2\bar{b}+ \sum_{j=1}^{n} {s_j^+ \over \lambda - \epsilon_j } \equiv 
 2\bar{b} {\prod_{k=1}^n(\lambda - \lambda_k) \over \prod_{j=1}^{n} (\lambda - \epsilon_j) }
 \label{Csepare}
\end{equation}
the separated coordinates are  the collection $(\lambda_k, \mu_k=A(\lambda_k))$. They have canonical Poisson brackets
\begin{equation}
\{ \lambda_k, \mu_{k'} \} = -i \delta_{k,k'}
\label{poisep}
\end{equation}
Notice however that if $\bar{b}=0$ theses variables are not well defined.

There are only $2n$ such coordinates which turn out to be invariant under the global $U(1)$ rotation
generated by $H_{n+1}$:
$$
b\to e^{i\theta} \; b,\quad \bar{b}\to e^{-i\theta} \; \bar{b}, \quad s_j^-\to e^{i\theta} s_j^-, \quad  s_j^+\to e^{-i\theta} s_j^+
$$

So  they describe  the {\em reduced} model obtained by fixing the value of $H_{n+1}$ and taking 
into consideration only the dynamical variables invariant under this $U(1)$ action. The initial dynamical model can be recovered by adding the phase of the oscillator coordinates $\bar{b},b$ to the separated variables.

We show now how to reconstruct  spin coordinates from the separated variables $(\lambda_k, \mu_k)$ and the
oscillator coordinates $(\bar{b},b)$. 
For $C(\lambda)$ we have eq.(\ref{Csepare}). It is a rational fraction of $\lambda$ which has simple poles at $\lambda=\epsilon_j$
whose residue is $s_{j}^{+}$.  
For $A(\lambda)$, we write:
$$
A(\lambda) = {P_{n+1}(\lambda) \over  \prod_{j=1}^n (\lambda - \epsilon_j) }
$$
The polynomial $P_{n+1}(\lambda)- 2\lambda \prod_{j=1}^n (\lambda - \epsilon_j)$ is of degree $n-1$, 
and we know its value at the $n$ points $\lambda_j$ because:
\begin{equation}
P_{n+1}(\lambda_j) = \mu_j \prod_{k=1}^n (\lambda_j - \epsilon_k) 
\label{eqpn}
\end{equation}
Therefore, we can write:
\begin{equation}
P_{n+1}(\lambda) = 2\lambda \prod_{j=1}^n (\lambda - \epsilon_j)
+ \sum_i (\mu_i-2\lambda_i) \prod_{k=1}^n (\lambda_i - \epsilon_k){ \prod_{ l \neq i} (\lambda - \lambda_l) \over 
\prod_{l \neq i} (\lambda_i - \lambda_l)}
\label{pn+1}
\end{equation}
Once $A(\lambda)$ and $C(\lambda)$ are known, 
 $B(\lambda)$  is determined by:
\begin{equation}
B(\lambda)= {Q_{2n+2}(\lambda) - P_{n+1}^2(\lambda) \over C(\lambda) \prod_k (\lambda - \epsilon_k)^2 }
={1\over 2\bar{b}} { Q_{2n+2}(\lambda) - P^2_{n+1}(\lambda) \over 
\prod_i (\lambda -\lambda_i) \prod_k (\lambda - \epsilon_k) }
\label{conj}
\end{equation}
The polynomial in the numerator is of degree $2n$, and moreover it is divisible by 
$\prod_i (\lambda -\lambda_i)$ because $P_{n+1}^2(\lambda_i) = Q_{2n+2}(\lambda_i)$, so we can write
\begin{equation}
B(\lambda) = 2 b {\prod_i (\lambda -\bar{\lambda}_i)  \over 
\prod_k (\lambda - \epsilon_k) }
\label{lambdabar}
\end{equation}

In this construction, the variables $\bar{\lambda}_i$ and the corresponding $\bar{\mu}_i$ are complicated functions of the $( \lambda_i, \mu_i)$.
The set of physical configurations (often refered to here as the {\em real slice}) 
is obtained by writing that the set $(\bar{\lambda_i}, \bar{\mu_i})$ is the {\em complex conjugate} of the set $(\lambda_i, \mu_i)$. This leads to a set of complicated relations whose solution is not known in general, but that we will solve in the soliton case.

\bigskip

The Hamiltonians $H_j$ of the  reduced model are obtained by writing that the points $(\lambda_k,\mu_k)$ belong to the spectral curve. We get a linear system of equations
$\sum_j {\cal B}_{kj} H_j = V_k$
where
$$
{\cal B}_{kj} ={1\over \lambda_k-\epsilon_j},\quad V_k = {1\over 2} \left(\mu_k^2 -\lambda_k^2 -4 H_{n+1} -\sum_j {s^2\over(\lambda_k-\epsilon_j)^2} \right)
$$
Its solution is 
$$
H_i = \sum_k ({\cal B}^{-1})_{ik} V_k, \quad ({\cal B}^{-1})_{jp} = {\prod_{l\neq p} (\epsilon_j -\lambda_l) \prod_i ( \lambda_p-\epsilon_i) \over \prod_{i\neq j} (\epsilon_j-\epsilon_i) \prod_{l\neq p} (\lambda_p-\lambda_l) }
$$
\bigskip
Using the Poisson bracket eq.(\ref{poisep}), the equation of motion of the variable $\lambda_k$  with respect to $H_i$  is then (no summation over $k$):
\begin{equation}
\partial_{t_i} \lambda_k = i \mu_k ({\cal B}^{-1})_{ik},
\label{motioni}
\end{equation}

\bigskip

We are interested in the system with prescribed  real values of the
conserved quantities, $H_1$,...,$H_{n+1}$,  i.e. we take as coordinates the $(\lambda_j, H_j)$ instead of the $(\lambda_j, \mu_j)$.  This amounts to fixing the Liouville torus we work with, or equivalently  the spectral
polynomial $Q_{2n+2}(\lambda)$. In this setting, the $\mu_j$ are determined  by the equations
$$
\prod_{k=1}^n (\lambda_j-\epsilon_k) \mu_j =( \pm)_j \sqrt{ Q_{2n+2}(\lambda_j) }
$$
The equations of motion eq.(\ref{motioni}) must be complemented by the equation of motion for the phase of $b$.

We consider now the level set containing the critical point eq.(\ref{static}).
It corresponds to  a maximally degenerate spectral curve:
\begin{equation}
Q_{2n+2}(\lambda) =4 \prod_{l=1}^{n+1} (\lambda - E_l)^2
\label{Qdegenerate}
\end{equation}
where the zeroes of $Q_{2n+2}(\lambda)$ are  the roots of the classical Bethe equation:
\begin{equation}
2E + \sum_{j=1}^{n} {s e_j\over E -\epsilon_j} =0
\label{Ei0}
\end{equation}
The  separated variables $\lambda_i, \mu_i$ now satisfy
\begin{equation}
\mu_i =2 (\pm)_i  {\prod_l (\lambda_i- E_l) \over \prod_{j} (\lambda_i-\epsilon_j)}
\label{mui}
\end{equation}
The choice of sign here plays a  crucial role in the description of the various strata of the level set. 

We first examine the case where a  $\lambda_i$ is frozen at a root $E_l$. 
Let us assume that $Q_{2n+2}(\lambda)$ has a {\em real } root at $\lambda=E$.
This means that $A^2(\lambda) + B(\lambda)C(\lambda)$ vanishes when $\lambda=E$.
But for real $\lambda$, $A(\lambda)$ is real and one has $C(\lambda)= \overline{B(\lambda)}$ 
so that $A(\lambda)$, $B(\lambda)$, $C(\lambda)$ must all vanish at $\lambda=E$.
In particular, recalling eq.(\ref{Csepare}), this means that one of the separated variables, 
say $\lambda_1$, is frozen at the value $E$, {\em provided} $\bar{b} b \neq 0$.
This implies also that $\lambda-E$ divides simultaneously $A(\lambda)$, $B(\lambda)$ and
$C(\lambda)$, and therefore $(\lambda-E)^{2}$ divides $Q_{2n+2}(\lambda)$. A  real
root of $Q_{2n+2}(\lambda)$ is necessarily a double root and one $\lambda_i$ must be frozen at $E$.

Nothing simple can be said  in the case $Q_{2n+2}(\lambda)$ has a simple {\em complex} root. 
So, let us assume that it has  a {\it double} complex root $E$. Of course the complex conjugate $\bar{E}$ is also a
double root. 
Since $\Lambda (\lambda) = A^2(\lambda) + B(\lambda) C(\lambda) $ we have
$$
\Lambda' (\lambda) = 2 A(\lambda) A'(\lambda) + B'(\lambda)C(\lambda) + B(\lambda) C'(\lambda)
$$
So, if $\Lambda (\lambda) $ has a double zero $E$ we have
\begin{eqnarray*}
0&=&A^2(E) + B(E) C(E) \\
0&=& 2 A(E) A'(E) + B'(E)C(E) + B(E) C'(E)
\end{eqnarray*}
In contrast to the real case, we cannot infer from these equations that $C(E)=0$.
But if $C(E)=0$ that is if one $\lambda_i$  freezes at $E$, the first equation implies $A(E)=0$ and the second equation implies $B(E) C'(E)=0$. Therefore if the zero
 of $C(\lambda)$ at $\lambda=E$ is 
simple, then necessarily $B(E)=0$. 
But $B(E)$ is the complex conjugate of  $C(\bar{E})$ which must therefore vanish {\em provided} $\bar{b} b \neq 0$.
From this we conclude that another separated variable $\lambda_k$ freezes at $\bar{E}$. 
In contrast to the real case, freezing is not compulsory, but it is the possibility to freeze the $\lambda_i$ by complex conjugated pairs 
that leads to the description of the real slice and the stratification of the level set.

\bigskip

Coming back to eq.(\ref{mui}), we see  from eqs.(\ref{eqpn},\ref{Qdegenerate}) that:
$$
P_{n+1}(\lambda_i) = (\pm)_i \sqrt{ Q_{2n+2}(\lambda_i)} =  2 (\pm)_i  \prod_l (\lambda_i - E_l), \quad P_{n+1}(\lambda) = 2(\lambda^{n+1} - \sigma_1(\epsilon) \lambda^n + \cdots )
$$
If we take the $+$ sign for all $i$, then obviously $P_{n+1}(\lambda) = 2 \prod_l (\lambda - E_l)$, 
and taking the residue at $\lambda=\epsilon_j$ in $P_{n+1}(\lambda)/\prod_i (\lambda-\epsilon_i)$, we find $s_j^z=se_j$. This is the static solution corresponding to the critical point.

\bigskip

To go beyond this trivial solution, we divide the $\lambda_i$ into three sets $\lambda_i^{+}\in {\cal E}^{+}$ and $\lambda_i^{-}\in {\cal E}^{-}$ depending on the sign in this formula, and $\lambda_i^{0}\in {\cal E}^{0}$  is the set of $\lambda_i$ frozen at some $E_l$.

We define  the polynomials:
$$
{\cal P}(\lambda,t) = \prod_j (\lambda - \lambda_j(t)) =  {\cal P}^{(+)}(\lambda)  {\cal P}^{(0)}(\lambda)  {\cal P}^{(-)}(\lambda)
$$
The equations of motion  eq.(\ref{motioni}) read in the soliton case
$$
 \partial_{t_i} \lambda_k = \sqrt{-1} \mu_k ({\cal B}^{-1})_{ik} = 2 \sqrt{-1} (\pm)_k { \prod_{l'}(\lambda_k-E_{l'}) 
\prod_{l\neq k} (\epsilon_i-\lambda_l) \over \prod_{j\neq i} (\epsilon_i-\epsilon_j) \prod_{l\neq k} (\lambda_k-\lambda_l) }
$$
hence, for $ E_l \notin {\cal E}^{0}$,
$$
{\sum_k}'   (\pm)_k { 1\over  \lambda_k -E_l}  \partial_{t_i} \lambda_k = -2 \sqrt{-1}{\prod_{l} (\epsilon_i-\lambda_l) \over  \prod_{j\neq i} (\epsilon_i-\epsilon_j)}
  {\sum_k}'    {\prod_{l'\neq l}'(\lambda_k-E_{l'}) 
 \over (\lambda_k-\epsilon_i) \prod_{l\neq k}' (\lambda_k-\lambda_l) }
$$
where ${\sum_k}' $ means that the frozen $\lambda_k \in {\cal E}^0$ are excluded. We can rewrite this equation as
\begin{eqnarray*}
\partial_{t_i} \log { {\cal P}^{(+)}(E_l) \over {\cal P}^{(-)}(E_l)} &=& -2 \sqrt{-1}{\prod_{l} (\epsilon_i-\lambda_l) \over  \prod_{j\neq i} (\epsilon_i-\epsilon_j)}
{\sum_k}' {\rm Res }_{\lambda_k} {\prod_{l'\neq l}(z-E_{l'}) \over (z-\epsilon_i) \prod_{l} (z-\lambda_l) } \\
&=& 2 \sqrt{-1}{\prod_{l} (\epsilon_i-\lambda_l) \over  \prod_{j\neq i} (\epsilon_i-\epsilon_j)}\left( {\rm Res }_{\infty} + {\rm Res }_{\epsilon_i}  \right)
 {\prod_{l'\neq l}'(z-E_{l'}) \over (z-\epsilon_i) \prod_{l}' (z-\lambda_l) } \\
 &=& 2 \sqrt{-1}{\prod_{l} (\epsilon_i-\lambda_l) \over  \prod_{j\neq i} (\epsilon_i-\epsilon_j)} \left(- 1+ {\prod_{l'\neq l}'(\epsilon_i-E_{l'}) \over \prod_{l}' (\epsilon_i-\lambda_l) }   \right)
\end{eqnarray*}
so that
$$
\partial_{t_i} \log { {\cal P}^{(+)}(E_l) \over {\cal P}^{(-)}(E_l)} = - 2 \sqrt{-1}{\prod_{l} (\epsilon_i-\lambda_l) \over  \prod_{j\neq i} (\epsilon_i-\epsilon_j)} 
+ 2 {\sqrt{-1}\over \epsilon_i-E_l}
{\prod_{l'}(\epsilon_i-E_{l'}) \over  \prod_{j\neq i} (\epsilon_i-\epsilon_j)}
$$
Remembering  the identities:
\begin{equation}
2 { \prod_k (\epsilon_j- E_k) \over \prod_{k\neq j} (\epsilon_j-\epsilon_k)} = s e_j, \quad 
\sum_l E_l  = \sigma_1(\epsilon)
\label{remember}
\end{equation}
we can rewrite
$$
\partial_{t_i} \log { {\cal P}^{(+)}(E_l) \over {\cal P}^{(-)}(E_l)} = -  \sqrt{-1}{s_i^{+}\over  \bar{b}} 
+  \sqrt{-1}{s e_i\over \epsilon_i-E_l}
$$
But the equation of motion for $\bar{b}$ is
$$
\partial_{t_i} \bar{b} = \{ H_i, \bar{b} \} = \sqrt{-1} s_i^+
$$
so that finally
$$
\partial_{t_i} \log { {\cal P}^{(+)}(E_l) \over {\cal P}^{(-)}(E_l)} =- \partial_{t_i} \log \bar{b} +  \sqrt{-1}{s e_i\over \epsilon_i-E_l}, \quad E_l \notin {\cal E}^{0}
$$
Integrating these equations we get
\begin{equation}
{\cal P}^{(-)}(E_l)  =  \bar{b}(\{ t \})  X_l(\{ t \})  {\cal P}^{(+)}(E_l), \quad X_l(\{ t \}) = X_l(0) e^{\sqrt{-1}\left( \sum_i {s e_i\over E_l-\epsilon_i} t_i 
- t_{n+1}\right)}, \quad E_l \notin {\cal E}^{0}
\label{cond1}
\end{equation}
 The $t_{n+1}$ dependence  of $X_l$ is fixed by  noticing that ${\cal P}^{(\pm)}(E_l)$ are independent of 
$t_{n+1}$, while $\partial_{t_{n+1}}\bar{b} = \sqrt{-1}  \;\bar{b}$.

Denoting $n_\pm,n_0$ the number of elements in ${\cal E}^{(\pm)}, {\cal E}^{0}$ respectively,
we get  a set of $n-n_0+1$ {\em linear} equations for the $n_+ + n_- + 1=n-n_0+1$ unknown coefficients of  
the polynomials ${\cal P}_{-}(\lambda)$ and $\bar{b}(t){\cal P}_{+}(\lambda)$. 

These real soliton solutions live on the pre-images of focus-focus singularities 
These are composed of various strata, depending on the 
number of $n_0$ of separated variables  frozen by pairs at the double roots $(E_l, E_{\bar{l}} = E_l^*)$ of the spectral polynomial. Separated variables  are necessarily frozen
at the real double roots of the spectral polynomial. These strata are themselves products of $(n+1-n_0)/2$ two dimensional pinched tori.

\bigskip

We then consider   the remaining complex roots of the spectral polynomial where no separated variables are frozen. This is a self conjugate set.
We write eq.(\ref{cond1})  as a linear system 
for the symmetric functions $\sigma^{(-)}_i(\{\lambda_j^{(-)} \} )$ and $\sigma^{(+)}_i(\{\lambda_j^{(+)} \} )$ and $\bar{b}$. We introduce the vector
$$
V= \pmatrix{(-1)^{n_-} \sigma_{n_-}^{(-)}/\bar{b} \cr \vdots \cr -\sigma_{1}^{(-)}/\bar{b} \cr 1/\bar{b} \cr (-1)^{1+n_+ } \sigma_{n_+}^{(+)} \cr \vdots \cr \sigma_1^{(+)} }
$$
then eq.(\ref{cond1}) reads
$$
(1,E,\cdots,E^{n_-},X,XE,\cdots, XE^{n_+-1} ) V = XE^{n_+}
$$
where we have defined the column vectors $(E^j)_l = E_l^j$  and $(XE^j)_l = X_lE_l^j$ where $X_l$ is given by eq.(\ref{cond1}). The formulae used in the text are straightforward consequences  Cramer's solution of this linear system.

In the vector $X$ we should incorporate the reality conditions
$$
 \overline{X_l} \; X_{\bar{l}} = -{1\over 4}, \quad E_{\bar{l}} \equiv \overline{E_l}
$$
which we derive below.

 \subsection{Reality conditions.}
 It remains to impose the condition that the set of zeros $\{\bar{\lambda}_k\}$ of $B(\lambda)$ in eq.(\ref{lambdabar}) is the complex conjugate of the 
 set of zeros $\{{\lambda}_k\}$ of $C(\lambda)$. Going back to eq.(\ref{conj}), 
the first  step is to build  the polynomial $P_{n+1}(\lambda)$.  Using the fact that $P_{n+1}(\lambda_i) = 2(\pm)_i \prod_l (\lambda_i- E_l)$ we may write : 
$$
P_{n+1}(\lambda) = 2 \prod_l  (\lambda -E_l) - 4 \sum_{\lambda_i^{-}} \prod_{E_l\notin {\cal E}^{0}} (\lambda_i^{-} -E_l) {\prod_{\lambda_l^{-} \neq \lambda_i^{-} } (\lambda- \lambda_l^{-}) \over \prod_{\lambda_l^{-} \neq \lambda_i^{-} }  ({\lambda_i^{-} - \lambda_l^{-} } )}
{{\cal P}_+(\lambda){\cal P}_0(\lambda) \over {\cal P}_+(\lambda_i^{-}) }
$$
To derive this formula, we used the fact that the polynomials $P_{n+1}(\lambda)$
and $2 \prod_l  (\lambda -E_l)$ have the same terms of degrees $n+1$ and $n$; the last
statement comes from the relation $\sum_{j}\epsilon_{j}=\sum_{l}E_{l}$ which is a direct consequence
of the classical Bethe equation. 
Alternatively, we can also write:
\begin{eqnarray*}
P_{n+1}(\lambda) &=& -2 \prod_l (\lambda -E_l) + 4 \sum_{\lambda_i^{+}} \prod_{E_l\notin {\cal E}^{0}} (\lambda_i^{+} -E_l) {\prod_{\lambda_l^{+} \neq \lambda_i^{+} } (\lambda- \lambda_l^{+}) \over \prod_{\lambda_l^{+} \neq \lambda_i^{+} }  ({\lambda_i^{+} - \lambda_l^{+} } )}
{{\cal P}_-(\lambda){\cal P}_0(\lambda) \over {\cal P}_-(\lambda_i^{+}) } \\
&&+4(\lambda + \Sigma_1' -\sigma_1'(E) ) {\cal P}_+(\lambda){\cal P}_-(\lambda){\cal P}_0(\lambda)
\end{eqnarray*}
Note that the last term in the right hand side is necessary to adjust the terms of degree $n+1$
and $n$ in $\lambda$ between the two sides of the equation. Again, we used the relation
$\sum_{j}\epsilon_{j}=\sum_{l}E_{l}$.
These two expressions for  $P_{n+1}(\lambda)$ motivate the following definitions of polynomials ${\cal S}_\pm(\lambda)$:
\begin{eqnarray*}
{\cal S}_+(\lambda) &=& \sum_{\lambda_i^{-}} \prod_{E_l\notin {\cal E}^{0}} (\lambda_i^{-} -E_l) {\prod_{\lambda_l^{-} \neq \lambda_i^{-} } (\lambda- \lambda_l^{-}) \over {\cal P}'_-(\lambda_i^{-}){\cal P}_+(\lambda_i^{-})  } \\
{\cal S}_-(\lambda) &=& \sum_{\lambda_i^{+}} \prod_{E_l\notin {\cal E}^{0}} (\lambda_i^{+} -E_l) {\prod_{\lambda_l^{+} \neq \lambda_i^{+} } (\lambda- \lambda_l^{+}) \over {\cal P}'_+(\lambda_i^{+})  {\cal P}_-(\lambda_i^{+})}
+(\lambda + \Sigma_1' -\sigma_1'(E) ) {\cal P}_+(\lambda)
\end{eqnarray*}
so that we can write:
\begin{eqnarray}
P_{n+1}(\lambda) &=& 2 \prod_l (\lambda -E_l) - 4 {\cal S}_+(\lambda) {\cal P}_+(\lambda) {\cal P}_0(\lambda) \label{Pn+1+} \\
P_{n+1}(\lambda) &=&- 2 \prod_l (\lambda -E_l) + 4 {\cal S}_-(\lambda) {\cal P}_-(\lambda)  {\cal P}_0(\lambda)\label{Pn+1-} 
\end{eqnarray}
Note that ${\cal S}_+(\lambda)$ has degree $n_{-}-1$ and  ${\cal S}_{-}(\lambda)$ has degree $n_{+}+1$.
Now, we have:
$$
4 \prod_l (\lambda -E_l)^2 - P_{n+1}^2(\lambda) =
16\; {\cal S}_-(\lambda) {\cal S}_+(\lambda)  {\cal P}_0(\lambda) {\cal P}(\lambda) = 4 \bar{b} b  {\cal P}(\lambda)  \bar{\cal P}(\lambda)
$$
where ${\cal P}(\lambda)=\prod_{j}(\lambda-\lambda_{j}) = {\cal P}_-(\lambda){\cal P}_0(\lambda){\cal P}_+(\lambda)$ and 
$\bar{\cal P}(\lambda)=\prod_{j}(\lambda-\bar{\lambda}_{j})$ is the complex conjugate of ${\cal P}(\lambda)$, that is
$\bar{\cal P}(\lambda) = \overline{ {\cal P}(\bar{\lambda})} $.  Therefore:
\begin{equation}
\bar{b} b \;  \bar{\cal P}(\lambda) =4\; {\cal S}_-(\lambda) {\cal S}_+(\lambda)  {\cal P}_0(\lambda)
\label{PbarSS1}
\end{equation}
So the zeroes $\bar{\lambda}_i$ of $\bar{\cal P}(\lambda)$ split into the zeroes 
$\bar{\lambda}_i^{+}$, $\bar{\lambda}_i^{-}$ and $\bar{\lambda}_i^{0}$ of 
$ {\cal S}_+(\lambda)$, $ {\cal S}_-(\lambda)$ and $ {\cal P}_0(\lambda)$ respectively. 
A direct consequence of these definitions and of eqs.~(\ref{Pn+1+}) and (\ref{Pn+1-}) is that:
\begin{eqnarray}
P_{n+1}(\bar{\lambda}_i^{+}) & = & + 2 \prod_l (\bar{\lambda}_i^{+} -E_l) \label{signPn+1bar+} \\
P_{n+1}(\bar{\lambda}_i^{-}) & = & - 2 \prod_l (\bar{\lambda}_i^{-} -E_l) \label{signPn+1bar-}
\end{eqnarray}

By the definition of the $\bar{\lambda}_i^{0}$'s, we see that 
the set ${\cal E}^{0}$ is self conjugate  and that ${\cal P}_0(\lambda)=\bar {\cal P}_0(\lambda)$.
The above definition of ${\cal S}_{-}(\lambda)$ shows that the coefficient of its term of highest
degree is equal to one (this is {\em not} the case for ${\cal S}_{+}(\lambda)$). Because of this:
\begin{equation}
{\cal S}_-(\lambda) = \bar{\cal P}_-(\lambda)
\label{Sm}
\end{equation}
Combining this with eq.~(\ref{PbarSS1}) we get also:
\begin{equation}
{\cal S}_+(\lambda) = {1 \over 4} \bar{b} b\; \bar{\cal P}_+(\lambda)
\label{Sp}
\end{equation}
Comparing the terms of highest degrees in ${\cal S}_+(\lambda)$ and $\bar{\cal P}_+(\lambda)$ gives: 
\begin{equation}
\bar{b} b = 4  \sum_{\lambda_i^{-}} { \prod_{E_l\notin {\cal E}^{0}} (\lambda_i^{-} -E_l) \over { {\cal P}'_-(\lambda_i^{-}){\cal P}_+(\lambda_i^{-})  }}
\label{bbar}
\end{equation}
It is expressed only in terms of $\lambda_i^{-}$'s. So, if $n_-=0$, we recover the fact already mentioned that $\bar{b} b =0$
and the system remains at the critical point.

At this stage, we are ready to enforce the reality condition. As discussed above,
the real slice is obtained by imposing that the set $\{\bar{\lambda}_{i}\}$
be the same as the set $\{\lambda_{i}^{*}\}$, where in the rest of this section we denote by $z^{*}$ the complex conjugate
of $z$. Equivalently: 
$$
\bar{\cal P}(\lambda^{*})={\cal P}(\lambda)^{*}
$$
for any $\lambda$. 
From the discussion at the beginning of this section, 
we know that the frozen variables
$\lambda_{i}^{0}$ appear in complex conjugate pairs so that 
${\cal P}_{0}(\lambda^{*})={\cal P}_{0}(\lambda)^{*}$.
We also know that  ${\cal P}_0(\lambda)=\bar {\cal P}_0(\lambda)$ so that 
$\bar{\cal P}_{0}(\lambda^{*})={\cal P}_{0}(\lambda)^{*}$.
The above reality condition becomes then:
$$
\bar{\cal P}_{-}(\lambda^{*})\bar{\cal P}_{+}(\lambda^{*})
={\cal P}_{-}(\lambda)^{*}{\cal P}_{+}(\lambda)^{*}
$$
It is clearly sufficient to impose simultaneously:
\begin{equation}
\label{necessary_sufficient}
\bar{\cal P}_{-}(\lambda^{*}) = {\cal P}_{-}(\lambda)^{*} ,\quad \bar{\cal P}_{+}(\lambda^{*}) = {\cal P}_{+}(\lambda)^{*}
\end{equation}
but we  claim that this condition is also necessary.
This comes from the fact already noted that the sign of $P_{n+1}(\lambda)/\prod_{l}(\lambda-E_{l})$ is positive
for $\lambda=\lambda_i^{+}$ or $\lambda=\bar{\lambda}_{i}^{+}$ and negative for $\lambda=\lambda_i^{-}$ or $\lambda=\bar{\lambda}_{i}^{-}$.
So the roots $\bar{\lambda}_i^{+}$ have to be complex conjugates of $\lambda_i^{+}$ and likewise,
the roots $\bar{\lambda}_i^{-}$ have to be complex conjugates of $\lambda_i^{-}$. An interesting and useful
consequence of this is that
we must have ${\rm deg } \;\;{\cal P}_\pm = {\rm deg }\;\; {\cal S}_\pm = {\rm deg }\;\; \bar{\cal P}_\pm$, which
requires $n_+=n_- -1$. Since $n_++n_0+n_-=n$ we find: 
$$
n_+={1\over 2}(n-1-n_0),  \quad n_-= {1\over 2} (n+1-n_0)
$$

Because the coefficients of highest degrees
of ${\cal P}_{-}(\lambda)$  and  ${\cal P}_{+}(\lambda)$ are set equal to one,
the above constraints~(\ref{necessary_sufficient}) are equivalent  to (assuming of course that $\bar{\cal P}_{+}(\lambda)$ and $\bar{\cal P}_{-}(\lambda)$ are mutually prime
and similarly for ${\cal P}_{+}(\lambda)$ and ${\cal P}_{-}(\lambda)$). 
$$
{\bar{\cal P}_{-}(\lambda^{*}) \over \bar{\cal P}_{+}(\lambda^{*})} =
\left({{\cal P}_{-}(\lambda) \over {\cal P}_{+}(\lambda)}\right)^{*} 
$$
We also note that we should add the constraint $\bar{b}=b^{*}$.
This plus the fact that these two polynomials involve a total of $n_{+}+n_{-}$ unknown coefficients,
shows that it is necessary and sufficient to enforce
the following conditions for the $n+1-n_{0}=n_{+}+n_{-}+1$ roots $E_{l}$ which don't belong to ${\cal E}^{0}$: 
\begin{equation}
{\bar{\cal P}_{-}(E_{l}^{*}) \over b\bar{\cal P}_{+}(E_{l}^{*})} =
\left({{\cal P}_{-}(E_{l}) \over \bar{b}{\cal P}_{+}(E_{l})}\right)^{*} 
\label{criterion_cal_P}
\end{equation}

As we have seen, the general solution of the Hamiltonian evolution on the critical torus, eq.~(\ref{cond1}) implies:
\begin{equation}
{{\cal P}_{-}(E_{l}) \over \bar{b}{\cal P}_{+}(E_{l})} = X_l 
\label{RHS_criterion}
\end{equation}
To evaluate the left-hand side of the conditions (\ref{criterion_cal_P}), we set $\lambda=E_{l}$
in eqs.(\ref{Pn+1+}, \ref{Pn+1-}), which gives:
$$
P_{n+1}(E_l) = -4 {\cal S}_+(E_l) {\cal P}_+(E_l) {\cal P}_0(E_l),\quad P_{n+1}(E_l) = 4 {\cal S}_-(E_l) {\cal P}_-(E_l) {\cal P}_0(E_l),
\quad E_l \notin {\cal E}^{0}
$$
and therefore:
\begin{equation}
 {\cal S}_+(E_l) {\cal P}_+(E_l) = - {\cal S}_-(E_l) {\cal P}_-(E_l)
 \label{cond2}
 \end{equation}
Since ${\cal P}_+(E_l) \neq 0$, and 
remembering eq.(\ref{cond1}) this implies: 
$$
{{\cal S}_{-}(E_l) \over {\cal S}_{+}(E_l)} = - {1 \over \bar{b} X_l}  
$$
But using eqs.~(\ref{Sm}) and (\ref{Sp}), we get:
\begin{equation}
{\bar{\cal P}_{-}(E_{l}^{*}) \over b\bar{\cal P}_{+}(E_{l}^{*})}=-{1\over 4 X_{\bar{l}}}
\label{LHS_criterion}
\end{equation}
where we define the index $\bar{l}$ to be such that $E_{\bar{l}}=E_{l}^{*}$.
Given eqs.~(\ref{RHS_criterion}) and (\ref{LHS_criterion}), the reality conditions (\ref{criterion_cal_P})
and the condition $\bar{b}=b^{*}$ are satisfied if and only if:
\begin{equation}
X_l^{*} \; X_{\bar{l}} = -{1\over 4}
\label{explicit_reality_conditions}
\end{equation}
As expected, times disappears from these conditions so that they reduce to constraints on the integration constants:
$$
X_l(0)^{*} \; X_{\bar{l}}(0) = -{1\over 4}   
$$
This characterizes a stratum of dimension $n-n_0+1$ on the real slice.

\section{Appendix : Asymptotic behavior of solitonic formulae}
\label{asymptotic}

Let us assume that the $m$ solitonic amplitudes $X_j$ become large while keeping fixed ratios. Because of the reality condition~(\ref{explicit_reality_conditions}),
the $m$ remaining amplitudes $X_{\bar{j}}$ go to zero with fixed ratios. With no loss of generality, let us assume that $j$ runs from 1 to $m$
and $\bar{j}$ from $m+1$ to $2m$. The complete expression~(\ref{Clambda}) for $C(\lambda)$ involves three determinants $D_1$, $D_2$ and $D$ defined by:
$$ 
D_1 = {\rm det} \pmatrix{ 1 & \lambda & \cdots & \lambda^{m} & 0 & 0     & \cdots &  0 \cr
                                                                                                      1 & E            & \cdots & E^{n_-}             & X & X E & \cdots & XE^{m-1} }
$$
$$
D_2 = {\rm det} \pmatrix{ 0 & 0 & \cdots & 0             & 1  & \lambda     & \cdots &  \lambda^{m-1} \cr
                                                                                                  1 & E  & \cdots & E^{m} & X & X E             & \cdots  & XE^{m-1}}
$$
$$
D= {\rm det} \pmatrix{  1& E& \cdots& E^{m-1}& X& XE& \cdots &  XE^{m-1} }
$$
In the limit where $|X_j|\rightarrow \infty$ and $X_{\bar{j}} \rightarrow 0$ for $1 \leq j \leq m$, the dominant term in $D_1$ is given by:
$$
D_1 \simeq {\rm det} \pmatrix{ 1 & \lambda & \cdots & \lambda^{m} & 0 & 0 & \cdots & 0 \cr
                    0 & 0 & \cdots & 0 & X_1 & X_1 E_1 & \cdots & X_1 E_1^{m-1} \cr
                    . & . &  & . & . & . &  & . \cr
                    . & . &  & . & . & . &  & . \cr
                    0 & 0 & \cdots & 0 & X_m & X_m E_m & \cdots & X_m E_m^{m-1} \cr
                    1 & E_{\bar{1}} & \cdots &  E_{\bar{1}}^{m} & 0 & 0 & \cdots & 0 \cr
                    . & . &  & . & . & . &  & . \cr
                    . & . &  & . & . & . &  & . \cr
                    1 & E_{\bar{m}} & \cdots &  E_{\bar{m}}^{m} & 0 & 0 & \cdots & 0}  
$$
So, we get:
\begin{equation}
D_1 \simeq \prod_{i<j}|E_j-E_i|^{2}\prod_{j=1}^{m}(\lambda-E_{\bar{j}}) \; X_1 \cdots  X_m
\label{D1_approx}
\end{equation}
It is also important to note that the subleading terms of $D_1$ vanish. We obtain such terms either by
removing one of the $X_j$'s or by adding one of the $X_{\bar{j}}$'s. In the first case, the $m$ last
columns to the right have only $m-1$ non-vanishing lines, so they are linearly dependent. Likewise,
in the second case, the first $m+1$ columns on the left have only $m$ non-vanishing lines. So the 
first corrections to $D_1$ originate from the terms where we simultaneously remove one $X_j$
and add one $X_{\bar{k}}$.

The leading term in $D_2$ has the form:
$$
{\rm det} \pmatrix{ 0 & 0 & \cdots & 0 & 1 & \lambda & \cdots & \lambda^{m-1} \cr 
                    0 & 0 & \cdots & 0 & X_1 & X_1 E_1 & \cdots & X_1 E_1^{m-1} \cr
                    . & . &  & . & . & . &  & . \cr
                    . & . &  & . & . & . &  & . \cr
                    0 & 0 & \cdots & 0 & X_m & X_m E_m & \cdots & X_m E_m^{m-1} \cr
                    1 & E_{\bar{1}} & \cdots &  E_{\bar{1}}^{m} & 0 & 0 & \cdots & 0 \cr
                    . & . &  & . & . & . &  & . \cr
                    . & . &  & . & . & . &  & . \cr
                    1 & E_{\bar{m}} & \cdots &  E_{\bar{m}}^{m} & 0 & 0 & \cdots & 0}
$$
which is equal to zero. One has therefore to consider subleading terms. They can be obtained by removing one
of the $X_j$'s or by adding one of the $X_{\bar{j}}$'s. In fact, only the first option leads to non-zero terms.
Therefore, we may write:
$$
D_2 \simeq \sum_{j=1}^{m} {\rm det} \pmatrix{ 0 & 0 & \cdots & 0 & 1 & \lambda & \cdots & \lambda^{m-1} \cr 
                    0 & 0 & \cdots & 0 & X_1 & X_1 E_1 & \cdots & X_1 E_1^{m-1} \cr
                    . & . &  & . & . & . &  & . \cr
                    . & . &  & . & . & . &  & . \cr
                    0 & 0 & \cdots & 0 & X_{j-1} & X_{j-1} E_{j-1} & \cdots & X_{j-1} E_{j-1}^{m-1} \cr
                    1 & E_j & \cdots & E_j^{m} & 0 & 0 & \cdots & 0 \cr
                    0 & 0 & \cdots & 0 & X_{j+1} & X_{j+1} E_{j+1} & \cdots & X_{j+1} E_{j+1}^{m-1} \cr
                     . & . &  & . & . & . &  & . \cr
                    . & . &  & . & . & . &  & . \cr
                    0 & 0 & \cdots & 0 & X_m & X_m E_m & \cdots & X_m E_m^{m-1} \cr
                    1 & E_{\bar{1}} & \cdots &  E_{\bar{1}}^{m} & 0 & 0 & \cdots & 0 \cr
                    . & . &  & . & . & . &  & . \cr
                    . & . &  & . & . & . &  & . \cr
                    1 & E_{\bar{m}} & \cdots &  E_{\bar{m}}^{m} & 0 & 0 & \cdots & 0}
$$
Explicitely:
\begin{equation}
D_2 \simeq \prod_{i<j}|E_j-E_i|^{2} \prod_{j=1}^{m}(\lambda-E_j) 
\sum_{j=1}^{m}\left(\frac{1}{X_j} \frac{E_{\bar{j}}-E_j}{\lambda-E_j}\prod_{k \neq j}\frac{E_{\bar{k}}-E_j}{E_k-E_j} 
\right) \; X_1 \cdots  X_m
\label{D2_approx}
\end{equation}
The same reasoning as for $D_1$ shows that the next corrections to this estimate of $D_2$ are obtained
by removing one more of the $X_j$'s and by adding simultaneously one of the $X_{\bar{k}}$'s. 

The determinant $D$ has a non-vanishing leading order term:
$$
D \simeq {\rm det} \pmatrix{ 0 & 0 & \cdots & 0 & X_1 & X_1 E_1 & \cdots & X_1 E_1^{m-1} \cr
                    . & . &  & . & . & . &  & . \cr
                    . & . &  & . & . & . &  & . \cr
                    0 & 0 & \cdots & 0 & X_m & X_m E_m & \cdots & X_m E_m^{m-1} \cr
                    1 & E_{\bar{1}} & \cdots &  E_{\bar{1}}^{m} & 0 & 0 & \cdots & 0 \cr
                    . & . &  & . & . & . &  & . \cr
                    . & . &  & . & . & . &  & . \cr
                    1 & E_{\bar{m}} & \cdots &  E_{\bar{m}}^{m} & 0 & 0 & \cdots & 0}
$$
so that:
\begin{equation}
D \simeq (-1)^{m} \prod_{i<j}|E_j-E_i|^{2}  \; X_1 \cdots  X_m
\label{D_approx}
\end{equation}

\section{Appendix: Periodic flows.}
\label{appendix_periodic}
There is a general procedure in the algebro-geometric setting to construct  Hamiltonians generating  periodic flows. From the equation of the spectral curve we have
$$
\delta \mu \; d \lambda = \sum_i {d\lambda \over \mu (\lambda-\epsilon_i)} \delta H_i = \sum_i \sigma_i(\lambda) d\lambda  \delta H_i
$$
where
$$
\sigma_i(\lambda) d\lambda = {d\lambda \over \mu (\lambda-\epsilon_i)}  = {\prod_{k\neq i} (\lambda-\epsilon_k) \over \sqrt{Q_{2n+2}(\lambda)}} d\lambda
$$
form a basis   of holomorphic differentials. Recalling the flows generated by $H_i$, eq.(\ref{motioni}), 
 we obtain the important relation
$$
\sigma_j(\lambda_k)\; \sum_k \partial_{t_i}\lambda_k   = i\sum_k  \mu_k ({\cal B}^{-1})_{ik} {{\cal B}_{kj} \over \mu_k} = i \delta_{ij}
$$
To find the Hamiltonian generating the periodic flows, we should need  the description of the periodic trajectory of the $n$ coordinates $\lambda_k$ of the divisor as $n$ ovals  on the spectral curve. 
However only the homotopy class of the trajectory will be important in the following discussion.

Let us assume for the time being that it is homotopic to $n$ independent non intersecting ovals $B_i$. Let us define 
$$
{\cal H}^{(i)} = \oint_{B_i} \sqrt{\Lambda(\lambda)} d\lambda
$$
To the $B_i$-cycles we associate a basis of normalized Abelian differentials $\omega_i(\lambda) d\lambda$. Of course we can expand it on the 
$\sigma_j(\lambda) d\lambda$ basis
$$
\omega_i (\lambda) d\lambda = \sum_j {\cal N}_{ij}\; \sigma_j(\lambda) d\lambda
$$
Equipped with these Abelian differentials, we define the angles through the Abel transformation
$$
\theta_j = \sum_k \int^{\lambda_k} \omega_j(\lambda) d\lambda
$$
Let us compute the equation of motion of these angles under the Hamiltonians ${\cal H}^{(i)}$.
\begin{eqnarray*}
\partial_{\tau_i} \theta_j &=& \sum_k \omega_j(\lambda_k) \partial_{\tau_i} \lambda_{k} =  \sum_{k,l} \omega_j(\lambda_k)  {\partial {\cal H}^{(i)} \over \partial H_l}\;
\partial_{t_l} \lambda_k  \\
&=& \sum_{l,n}  {\partial {\cal H}^{(i)} \over \partial H_l}\; {\cal N}_{jn} \sum_k \sigma_n(\lambda_k)\; \partial_{t_l} \lambda_k \\
&=& i \sum_{l,n}  {\partial {\cal H}^{(i)} \over \partial H_l}\; {\cal N}_{jn} \delta_{ln} = \oint_{B_i}  {\cal N}_{jl}{ \partial \over \partial H_l} \sqrt{\Lambda} d\lambda \\
&=& i \oint_{B_i}  {\cal N}_{jl} \; \sigma_l(\lambda) d\lambda =  i \oint_{B_i}  \omega_j(\lambda) d\lambda = 2i\pi \delta_{ij}
\end{eqnarray*}
Hence
$$
\partial_{\tau_i} \theta_j  = 2i\pi \delta_{ij}
$$
and the flows of the Hamiltonians ${\cal H}^{(i)}$ are $2\pi$-periodic. So the problem of finding the Hamiltonians generating the periodic flows reduces to the description of the homotopy class of the  $n$ non intersecting ovals of the periodic trajectories.

\end{document}